\documentclass[compsoc,conference,a4paper,10pt,times]{IEEEtran}
\IEEEoverridecommandlockouts
\usepackage{cite}
\usepackage{amsmath,amssymb,amsfonts}
\usepackage{algorithmic}
\usepackage{graphicx}
\usepackage{textcomp}
\usepackage{bmpsize}
\usepackage{xcolor}
\usepackage{enumitem}
\usepackage{lipsum}
\usepackage[colorlinks=true,citecolor=ForestGreen,urlcolor=ForestGreen,linkcolor=BrickRed]{hyperref}
\def\BibTeX{{\rm B\kern-.05em{\sc i\kern-.025em b}\kern-.08em
    T\kern-.1667em\lower.7ex\hbox{E}\kern-.125emX}}

\usepackage{xspace}
\usepackage[table,xcdraw,dvipsnames]{xcolor}

\usepackage{colortbl} 
\usepackage{multirow}
\usepackage{pifont}
\usepackage{booktabs}
\usepackage{adjustbox}
\usepackage{xstring} 
\usepackage{tikz}
\usepackage{xurl}  

\usepackage{amsmath,amssymb}
\usepackage[capitalize,nameinlink]{cleveref}
\crefname{section}{Sec.}{Secs.}
\Crefname{section}{Section}{Sections}

\crefname{subsection}{Sec.}{Secs.}
\Crefname{subsection}{Section}{Sections}

\crefname{figure}{Fig.}{Figs.}
\Crefname{figure}{Figure}{Figures}

\crefname{table}{Tab.}{Tabs.}
\Crefname{table}{Table}{Tables}
\usepackage{makecell}
\usepackage{subcaption}
\usepackage{booktabs}
\usepackage{pifont}
\newcommand{\cmark}{\ding{51}}% ✓
\newcommand{\xmark}{\ding{55}}% ✗
\newcommand{\MakeMarkups}[3]{
  \expandafter\newcommand\csname #2\endcsname[1]{\textcolor{#3}{\textbf{[##1]}}}
}

\definecolor{darkgreen}{rgb}{0.4, 0.6, 0.2}
\MakeMarkups{Sina}{SA}{blue}
\MakeMarkups{Amir}{AM}{purple}
\MakeMarkups{Hamed}{HH}{magenta}
\MakeMarkups{Marios}{MK}{cyan}

\usepackage{listings}
\usepackage{xcolor}
\definecolor{clKeyword}{HTML}{0057B8}  % deep blue
\definecolor{clComment}{HTML}{5C7E60}  % desaturated green
\definecolor{clString}{HTML}{B04300}   % rust
\definecolor{clBG}{gray}{0.97}
\definecolor{clFrame}{gray}{0.65}

\lstdefinestyle{CCompact}{
  language=C,
  columns=fixed,            % allow breaks at underscores
  keepspaces=true,
  basicstyle=\ttfamily\scriptsize,     % compact font
  keywordstyle=\color{clKeyword}\bfseries,
  commentstyle=\itshape\color{clComment},
  stringstyle=\color{clString},
  numbers=left,
  numberstyle=\tiny\color{clFrame},
  stepnumber=1,
  xleftmargin=1.25em,       % slim margins
  frame=single,
  framerule=0.3pt,
  rulecolor=\color{clFrame},
  backgroundcolor=\color{clBG},
  breaklines=true,
  breakindent=0pt,
  prebreak=\mbox{\textcolor{clFrame}{$\hookleftarrow$}},
  postbreak=\mbox{\textcolor{clFrame}{$\hookrightarrow$}},
  showstringspaces=false,
  aboveskip=4pt,
  belowskip=4pt,
  lineskip=1.2pt,
  tabsize=4
}

\usepackage{tikz}

\newcommand\eg{\emph{e.g.},\xspace}
\newcommand\ie{\emph{i.e.},\xspace}

\providecommand{\etal}{\emph{et al.}\xspace}

\newcommand{\para}[1]{\smallskip \noindent \textbf{#1.}}
\newcommand*\circled[1]{\tikz[baseline=(char.base)]{
            \node[shape=circle,draw, inner sep=0pt,minimum size=11pt] (char) {\footnotesize#1};}}

\newcommand{\name}{\textit{CAEC}\xspace}
\newcommand{\names}{\textit{CAEC}'s\xspace}

\begin{document}
\author{
\IEEEauthorblockN{
Sina Abdollahi$^{*}$,
Amir Al Sadi$^{*}$,
David Kotz$^{\dagger\ddagger}$\thanks{$^\ddagger$This work was performed while Professor Kotz was in residence at Imperial College London.},
Marios Kogias$^{*}$,
Hamed Haddadi$^{*}$}
\IEEEauthorblockA{
$^{*}$\textit{Imperial College London}, London, United Kingdom \\
\{s.abdollahi22, a.al-sadi, m.kogias, h.haddadi\}@imperial.ac.uk\\[0.5ex]
$^{\dagger}$\textit{Dartmouth College}, Hanover, NH, USA \\
David.F.Kotz@dartmouth.edu}
}

 \title{\name: {\underline C}onfidential, {\underline A}ttestable, and {\underline E}fficient Inter‑CVM \\ {\underline C}ommunication with Arm CCA\thanks{This paper has been accepted to appear in 11th IEEE European Symposium on Security and Privacy (EuroS\&P 2026)}}

 % \title{\name: {\underline C}onfidential, {\underline A}ttestable, and {\underline E}fficient Inter‑CVM \\ {\underline C}ommunication with Arm CCA}

% \title{Confidential, Attestable, and Efficient Inter‑CVM \\ Communication with Arm CCA\thanks{This paper has been accepted to appear in 11th IEEE European Symposium on Security and Privacy (EuroS\&P 2026)}}
\maketitle

\begin{abstract}
Confidential Virtual Machines (CVMs) are increasingly adopted to protect sensitive workloads from privileged adversaries such as the hypervisor. While they provide strong isolation guarantees, existing CVM architectures lack first-class mechanisms for inter-CVM data sharing due to their disjoint memory model, making inter-CVM data exchange a performance bottleneck in compartmentalized or collaborative multi-CVM systems. Under this model, a CVM’s accessible memory is either shared with the hypervisor or protected from both the hypervisor and all other CVMs. This design simplifies reasoning about memory ownership; however, it fundamentally precludes plaintext data sharing between CVMs because all inter-CVM communication must pass through hypervisor-accessible memory, requiring costly encryption and decryption to preserve confidentiality and integrity.
In this paper, we introduce \name, a system that enables protected memory sharing between CVMs. \name builds on  Arm Confidential Compute Architecture (CCA) and extends its firmware to support Confidential Shared Memory (CSM), a memory region securely shared between multiple CVMs while remaining inaccessible to the hypervisor and all non-participating CVMs. \names design is fully compatible with CCA hardware and introduces only a modest increase (6\%) in CCA firmware code size.
\name delivers substantial performance benefits across a range of workloads. For instance, inter-CVM communication over \name achieves up to  209$\times$ reduction in CPU cycles compared to encryption-based mechanisms over hypervisor-accessible shared memory.
By combining high performance, strong isolation guarantees, and attestable sharing semantics, \name provides a practical and scalable foundation for the next generation of trusted multi-CVM services across both edge and cloud environments.
\end{abstract}

\section{Introduction}
% tees, enclave and confidential VMs
Trusted Execution Environments (TEEs) have emerged as a cornerstone for building secure systems in the presence of powerful adversaries.
By providing isolated execution, integrity guarantees, and confidentiality for code and data even against a compromised operating system or hypervisor, TEEs enable developers to deploy sensitive applications on untrusted infrastructure.

Although initially the TEE landscape was rather heterogeneous with different vendors offering different mechanisms, \eg Intel providing enclaves with SGX~\cite{intelsgx} and Arm  offering physical memory partitioning with TrustZone~\cite{trustzone}, over the years there is a convergence towards confidential virtual machines (CVMs) as the prevalent abstraction for confidential computing.
Currently, major vendors have developed CVM offerings.
Technologies such as AMD SEV-SNP (Secure Encrypted Virtualization-Secure Nested Paging)~\cite{amdsevsnp} and Intel TDX (Trust Domain Extensions)~\cite{inteltdx} have already been adopted by cloud providers~\cite{edgelesssystem,awsllm,googleprivatespace,applepcc,CVMazure}, while Arm has similarly introduced Confidential Compute Architecture (CCA)~\cite{ccasite,armccaenablement}, which is expected to be deployed across both edge devices and cloud servers.

% the tee model of dijoint memory and its benefits
Despite design variations across vendors, all existing CVM architectures adopt a \textit{disjoint memory model}, in which each CVM owns a protected memory isolated from both the hypervisor and other CVMs~\cite{amdsevsnp, inteltdx, armccaenablement}. While each CVM may share memory with the hypervisor (\eg for I/O), it cannot create protected memory regions shared with other CVMs only and not visible to the hypervisor.
This design simplifies reasoning about memory ownership and provides strong confidentiality and integrity guarantees for the CVM’s protected memory region.
%

\begin{figure*}[t]
    \centering
    \includegraphics[width=0.95\linewidth]{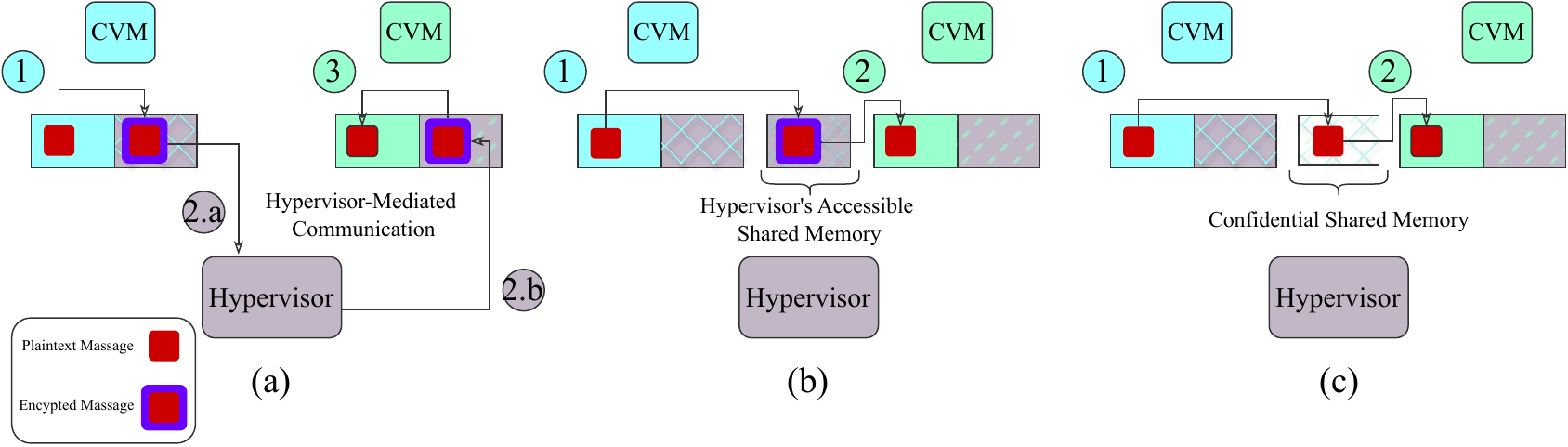}
   \caption{Communication modes between two CVMs. 
(a) Communication through virtualization services, where data must be encrypted and passed via the hypervisor-mediated service. 
(b) Communication using shared memory provided by the hypervisor, still requiring encryption as the shared memory is accessible to the hypervisor. 
(c) \name, which enables CSM between CVMs. There is no need for encryption as \name protects the CSM from hypervisor and other CVMs. Cyan: memory regions accessible to the first CVM, Green: memory regions accessible to the second CVM, Purple: memory regions accessible to the hypervisor.}       
    \label{fig:introduction}
\end{figure*}

% the problems with the disjoint model
Such a disjoint memory model though, leads to substantial efficiency losses in CVMs.
In practice, this design forces all inter-CVM communication to be exposed to the hypervisor, either through hypervisor-managed services such as virtual sockets (mode~(a) in \cref{fig:introduction}) or through hypervisor-accessible shared memory regions (mode~(b) in \cref{fig:introduction}).
Both approaches require expensive encryption and decryption on the CVM sides to preserve confidentiality and integrity when data is transmitted through hypervisor-accessible memory, while the use of confidential paravirtual devices is an open challenge both in terms of performance and correctness~\cite{misono2024confidential,li2023bifrost,abdollahi2025early, lefeuvre2023towards}. 
Furthermore, the disjoint memory model prevents memory savings through deduplication, where identical pages are shared across CVMs.
This inefficiency becomes particularly problematic not only in large-scale datacenter environments where DRAM is a major cost factor, but also in memory-constrained edge and embedded devices, where TEEs are increasingly deployed.

% why is it a timely problem
Unfortunately, communication and memory inefficiencies substantially affect modern machine learning (ML) and agentic systems, which increasingly dominate confidential computing workloads~\cite{edgelesssystem,awsllm,googleprivatespace,applepcc,CVMazure}.
These systems are increasingly becoming compartmentalized across CVMs, for example, splitting the networking and inference stacks into different components to reduce the attack surface~\cite{awsllm,edgelesssystem,nvidiaedgeless}. Moreover, ML and agentic systems provide large opportunities for memory deduplication and sharing between CVMs given the overlap in the models and contexts used across different applications, which can be multiple gigabytes.

% what is the solution
%\MK{Consider removing this paragraph, given that it is repetitive with the last part of the intro where you explain the contribution.}
%\SA{We define CSM here for the first time as a memory type. Later in contribuiton we say that we enabled this type of memory in CCA and protect it appropriately}
To address  this limitation, in this paper, we introduce Confidential Shared Memory (CSM), a  hypervisor-protected (confidential) memory which can be shared between multiple CVMs.
Designing a system that (1) enables CSM regions within a confidential computing environment and (2) restricts their use exclusively to mutually attested realms provides  CVMs with a protected memory for exchanging plaintext data directly with each other (mode (c) in \cref{fig:introduction}).
%\MK{It cannot be a channel. You are also talking about deduplication.}
This capability substantially improves CPU efficiency and memory utilization of inter-CVM data exchange.
%
%While the traditional disjoint memory model of CVMs is based on the distrust between workloads running , we envision scenarios where CVMs do not trust each other 
%
%

Enabling CSM may appear straightforward, yet it is a challenging undertaking.
Depending on the architecture, it may be infeasible without hardware modifications, for example, in AMD SEV-SNP (\cref{sec:enablingcsminother}).
%or may require microcode updates (\eg Intel TDX), both of which are difficult to develop and evaluate in practice (see 
%
Even systems that enable similar memory sharing between enclaves (user-space TEEs) fail to provide flexible and scalable mechanisms that fit the needs of modern ML systems.
For instance, Plug-In Enclaves~\cite{li2021confidential} enable read-only shared enclaves, which can be used to effectively share static resources such as libraries in serverless applications.
Cerberus~\cite{lee2022cerberus} introduces formal techniques to verify such a sharing with the same read-only access model. However, both  designs are insufficient for applications that require extensive inter-process communication (IPC), where writes must be visible to the other side.
Elasticlave~\cite{yu2022elasticlave}, on the other hand, supports writable shared memory using RISC-V Physical Memory Protection (PMP).
However, the underlying isolation mechanism inherently limits the number of shareable regions: the total number of memory partitions (protected or shared) is limited to the number of PMP registers (\eg to 16), thus, restricting both the number of coexisting enclaves and the number of shared regions. 

Even after extending a confidential computing architecture to support CSM across multiple CVMs, protecting the shared memory from attacks by other CVMs remains challenging. Participating CVMs must be able to discover and attest each other while being assured that the shared region cannot be accessed or modified by unauthorized CVMs. Without a principled ownership model and explicit access-control mechanisms, the system risks undermining the confidentiality of the CSM, enabling adversarial CVMs to exploit the CSM management interface to escalate their privileges or gain unauthorized access to shared regions.

\para{Contribution} 
In this paper, we introduce \name, a system that enables CSM between CVMs in Arm CCA. To the best of our knowledge, \name is the first approach to support CSM between CVMs. \name leverages Arm CCA's RISC architecture to enable CSM without requiring any modifications to the CCA hardware. It provides dynamically manageable memory sharing between CVMs without imposing any limitation on the number of shared regions, either per CVM or system-wide. Through these capabilities, \name enables efficient plaintext communication and the direct sharing of large resources (\eg, ML models) between mutually attested CVMs.
We implement \name on the latest functional and performance prototypes of Arm CCA. Our evaluation shows that \name achieves substantial performance improvements in inter-CVM communication and memory sharing such as up to 209$\times$ reduction in CPU cycles during communication.
% Summary of contributions
In summary, we claim the following contributions:
\begin{itemize}
  \item \name extends the CCA firmware to support the CSM. It introduces a principled ownership model, explicit access-control rules, and attestation extensions that ensure CSM remains protected from the hypervisor and all unauthorized CVMs. Our analysis shows that \name prevents unauthorized access to CSM while preserving CCA’s security guarantees for all non-CSM regions.
 \item \name integrates cleanly across all layers of the CCA stack, providing flexible CSM management capabilities to realms while retaining the hypervisor’s authority over physical memory management. \name achieves this with only a 4\% increase in firmware size.
  \item \name demonstrates significant benefit for communication and data sharing between CVMs. \name achieves  up to a 209$\times$
reduction in CPU cycles compared to encryption-based
mechanisms over hypervisor-accessible shared memory. Moreover using \name, big data like a LLM can be shared between two CVMs, resulting in  16.6\%--44.4\% reduction in the system's memory footprint\footnote{Our evaluation artifacts is available at \url{https://caec-paper.github.io/}}. 
 %  \item \name extends the attestation token of realms, providing a verifiable proof of the CSM enablement in each CVM.
   % \item All code of \name will be open-sourced upon acceptance.
\end{itemize}
\section{Background \& Motivation}
\begin{table}[tb]
  \centering
\caption{Memory access control applied by Granule Protection Check (GPC) in CCA}
      \resizebox*{\linewidth}{!}{%
    \begin{tabular}{l c c c c}
          \cline{1-5}
        \textbf{Security State} & \textbf{Normal PAS} & \textbf{Secure PAS} & \textbf{Realm PAS} & \textbf{Root PAS} \\
        \cline{1-5}
       Normal  & \cmark & \xmark  & \xmark  & \xmark \\
          Secure & \cmark & \cmark & \xmark & \xmark\\
         Realm & \cmark & \xmark &  \cmark & \xmark \\
         Root & \cmark & \cmark & \cmark & \cmark \\
       \cline{1-5}
    \end{tabular}
    }
    \label{tab:memoryaccess}
\end{table}
In this section, we review the foundational concepts required for the remainder of the paper. We begin with an overview of Arm CCA (\cref{sec:backgroundarmcca}), followed by a discussion of two key components of the architecture: the Realm Management Monitor (\cref{sec:rmm}) and the attestation framework (\cref{sec:attestation}). We conclude this section with the motivation behind this work (\cref{sec:motivation}).
%%%%%%%%%%%%%%%%%%%%%%%%%%%%%%%%%%%%%%%%%%%%%%%%%%%%%%%%%%%%%%%%%%%%%%
\subsection{Arm CCA}\label{sec:backgroundarmcca}
\begin{figure}[t]
    \centering
    \includegraphics[width=1\linewidth]{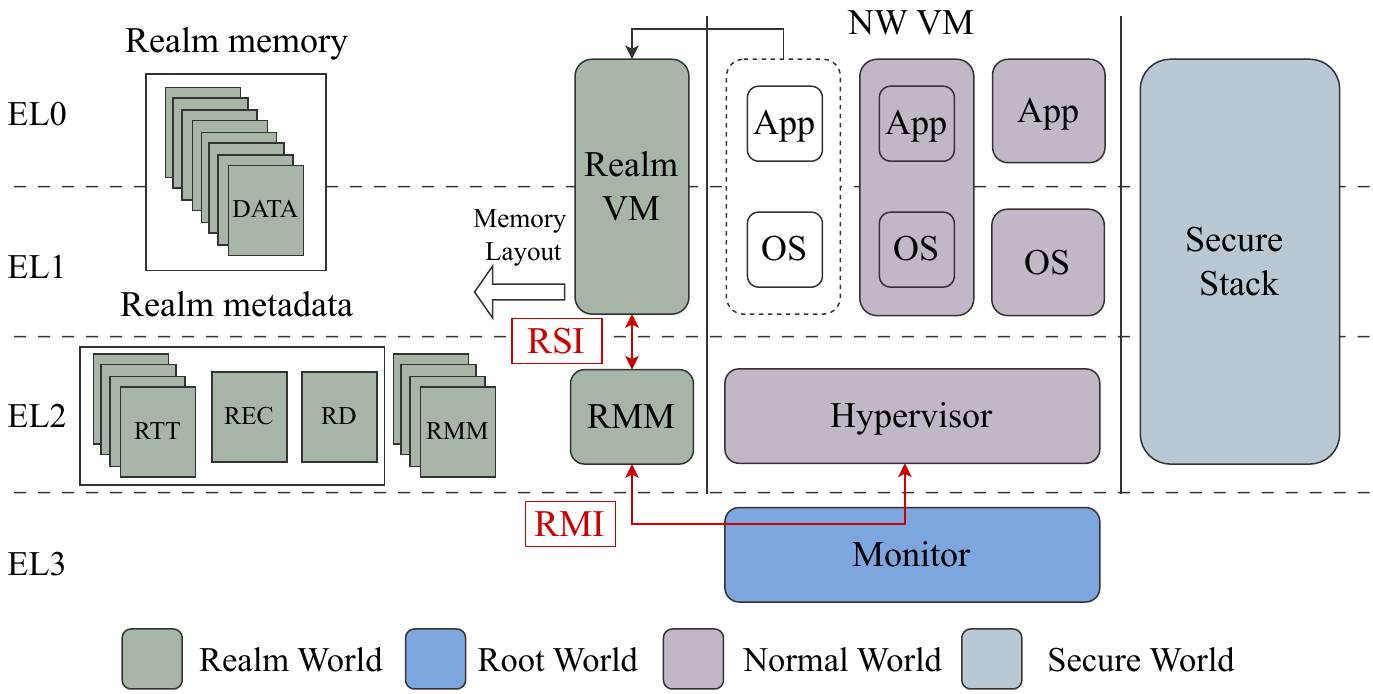}
   \caption{Arm CCA 1.0 software architecture. The hypervisor allocates resources (\eg memory and CPU) for realm VM but cannot access those resources, as the realm VM is running on the other side of the isolation boundaries.}
    \label{fig:Arm CCA software architecture}
\end{figure}

Arm CCA \cite{ccasite} is a series of hardware and software extensions for Armv9-A architecture (\cref{fig:Arm CCA software architecture}).  
Arm CCA extends the Armv9-A architecture with realm world\footnote{In some references, the term execution environment is used instead of world; however, in this work, the two terms are used interchangeably.} and  root world, orthogonal to the existing normal world (NW) and secure world.
Each 4~KB frame of physical memory (also referred to as a granule) is tagged with the world it belongs to at any given time.
This ownership information is recorded in a structure called Granule Protection Table (GPT).
A hardware mechanism called Granule Protection Check (GPC) enforces access restrictions based on both the ownership state of each granule (as recorded in the GPT) and the current processor state.
A memory access is permitted only if it complies with the rules defined in \cref{tab:memoryaccess}.
In particular, when the processor operates in the root world state, it has access to the physical address space (PAS) of all other worlds.
When operating in the realm or secure world, the processor can access the normal world’s PAS, but the realm and secure worlds are isolated from each other.
The normal world has no access to the PAS of any other world.
Arm CCA also leverages the isolation primitives of the Arm architecture—such as exception levels (ELs) and virtualization.
Exception levels begin with EL3, the highest privilege level in the system, while EL2–EL0 provide intra-world privilege separation.
The architecture support for virtualization includes two stages of address translation: mapping virtual addresses (VAs) to intermediate physical addresses\footnote{We adopt Arm’s term intermediate physical address (IPA), which is also commonly referred to as guest physical address (GPA).} (IPAs) and mapping IPAs to physical addresses (PAs).

Within CCA, the architecture software stack includes the Monitor running at EL3, responsible for initially booting all EL2 components, managing GPTs, and context switching between worlds. 
The normal world stack consists of a hypervisor operating at EL2, virtual machines (VMs) running at EL1 and EL0, and user-space apps running at EL0. Secure world can host a stack similar to the NW, however, it is usually reserved for vendor specific services, impossible to run third party code. The realm world stack consists of realm VMs (or simply realms) running at EL1 and EL0 and a lightweight firmware known as Realm Management Monitor (RMM). 
In CCA, the hypervisor retains control over system resources such as CPU cores and physical memory.
It can allocate or reclaim resources for realms much like it does for NW VMs.
However, in the case of realms, the RMM acts as a trusted mediator between the hypervisor and realms. It validates all hypervisor requests concerning realm resources and proceeds only if they satisfy CCA’s isolation and security requirements.
%Certain memory regions are carved out at boot time and reserved for critical components, including the RMM, the Monitor, and the Secure World stack.
\subsection{Realm Management Monitor} \label{sec:rmm}
The RMM is the trusted firmware component in CCA responsible for coordinating all interactions between the hypervisor and realms.
It enforces \textit{isolation} by protecting resources delegated to realms from both the hypervisor and other realms, while maintaining the hypervisor’s capability to \textit{manage} those resources.
To achieve this, the RMM exposes Realm Management Interface (RMI) to the hypervisor and the Realm Service Interface (RSI) to realms~\cite{rmmspec}.
The hypervisor can issue RMI commands to request operations on realm resources such as memory delegation or vCPU (virtual CPU) scheduling. However, before executing these operations, the RMM performs a series of validity checks and proceeds only if all validations succeed.
The RSI also can be used by realms to access services such as attestation and hypercalls.
The RMM manages four types of granules during the lifecycle of each realm (\cref{fig:Arm CCA software architecture}):
(1) the Realm Descriptor (RD), which defines the realm’s general attributes (\eg address space size);
(2) the Realm Execution Context (REC), which stores vCPU-related state (\eg system registers);
(3) the Realm Translation Tables (RTTs), a hierarchical structure maintaining IPA-to-PA mappings and access permissions; and
(4) DATA granules, which represent protected memory regions accessible only to realm software.

To allocate a new DATA granule to realm during runtime,
the hypervisor must first delegate a granule to the realm world using \texttt{RMI\allowbreak\_GRANULE\allowbreak\_DELEGATE}.
It then requests the RMM to create the corresponding mapping in the realm’s protected address space\footnote{A realm’s address space consists of two halves: the protected half, used to map realm-world granules, and the unprotected half, used to map normal-world granules.} using \texttt{RMI\_DATA\allowbreak\_CREATE\allowbreak\_UNKNOWN}.
Upon receiving the request, the RMM verifies that the granule has already been delegated to the realm world (\ie inaccessible to the hypervisor) and that it is not already mapped in the protected address space of another realm, thereby enforcing CCA’s disjoint memory model. After creating the new mapping in the RTT, the granule becomes accessible to realm.
%It is important to note that the RTTs of each realm must be populated separately using \texttt{RMI\_RTT\_CREATE} command. Mapping destruction is supported via \texttt{RMI\_DATA\_DESTROY} and granule undelegation is handled through \texttt{RMI\_GRANULE\_UNDELEGATE}.    
%
\subsection{Attestation}\label{sec:attestation}
In CCA, a realm can obtain an \textit{attestation token} via  RSI. The attestation token is a set of claim values and their signature that describe the state of the realm and the platform on which it runs~\cite{rmmspec,sardar2023sok}.
A token includes claims such as the Realm Initial Measurement (RIM), which captures the realm’s configuration and initial memory contents. Because the RMM computes these claims and signs them as part of the token, a remote verifier (\eg a realm owner) can (1) validate their authenticity, ensuring that each claim was generated by the trusted RMM and has not been tampered with and (2) check whether these claims match the expected values.
%Thus,  Therefore, adding new claims to the RMM

\subsection{Motivation}\label{sec:motivation}
The traditional disjoint memory model of CVMs forbids any form of CSM between CVMs. This design is reasonable under the common assumption that CVMs do not trust one another. However, it is too restrictive for emerging workloads, which increasingly require CVMs—while still isolated from each other—to collaborate once attestation confirms that a peer is running an expected and acceptable software stack. These collaborating CVMs may belong to different administrative parties (\eg agentic systems~\cite{wu2024isolategpt} or collaborative learning~\cite{chen2024protecting}), or they may belong to a single organization but be compartmentalized for security (\eg separating networking and  inference stack~\cite{awsllm}). Under the current trust model, such workloads cannot efficiently exchange plaintext data or share large identical memory regions (\eg LLM model weights), even though the shared content itself poses no confidentiality risk.
Meeting these emerging demands requires rethinking the architecture of confidential computing: CVMs must be able to share CSM with attested peers while remaining fully protected from untrusted or malicious CVMs.
%
%For instance, a secure ML pipeline may split networking and inference across different CVMs to reduce the attack surface. Similarly, multiple CVMs might reuse the same pre-trained model or context for their own services.

\para{Arm CCA for CSM}
Arm CCA introduces architectural properties that make it a promising candidate for supporting flexible and secure CSM. First, unlike AMD SEV-SNP or Intel TDX, which rely extensively on hardware and microcode extensions, CCA places most of its trusted functionality in firmware (\ie the RMM) while relying on a small set of hardware mechanisms to enforce isolation~\cite{li2022design,ccasoftwarearchitecture}. This design provides better flexibility for extending the architecture with new features. For example, CCA uses the GPC hardware mechanism to isolate the realm world from the normal world, but isolation between realms is enforced entirely through RMM-managed checks. Because these checks reside in firmware rather than hardware, they can be extended or relaxed through firmware updates. Similarly, the RMM can incorporate new metadata types and validation rules without requiring hardware or microcode modifications—capabilities that are essential for securely managing CSM regions across realms.
Second, in contrast to the physical partitioning mechanism used in Elasticlave~\cite{yu2022elasticlave}, CCA’s use of virtualization can be adapted to support an unbounded number of regions, both within a single CVM and across the system, albeit with increased design complexity.
Overall, these properties suggest that CCA provides a solid foundation for implementing a scalable and fine-grained CSM mechanism within a confidential computing architecture.
%%%%%%%%%%%%%%%%%%%%%%%%%%%%%%%%%%%%%%%%%%%%%%%%%%%%%%%%%%%%%%%%%%%%%%
\section{System \& Threat Model}
In this section, we present the system model (\cref{sec:systemmodel}) and the threat model (\cref{sec:adversarymodel}) of \name.
%%%%%%%%%%%%%%%%%%%%%%%%%%%%%%%%%%%%%%%%%%%%%%%%%%%%%%%%%%%%%%%%%%%%%%
\subsection{System Model}\label{sec:systemmodel}
\begin{figure}[t]
    \centering
    \includegraphics[width=1\linewidth]{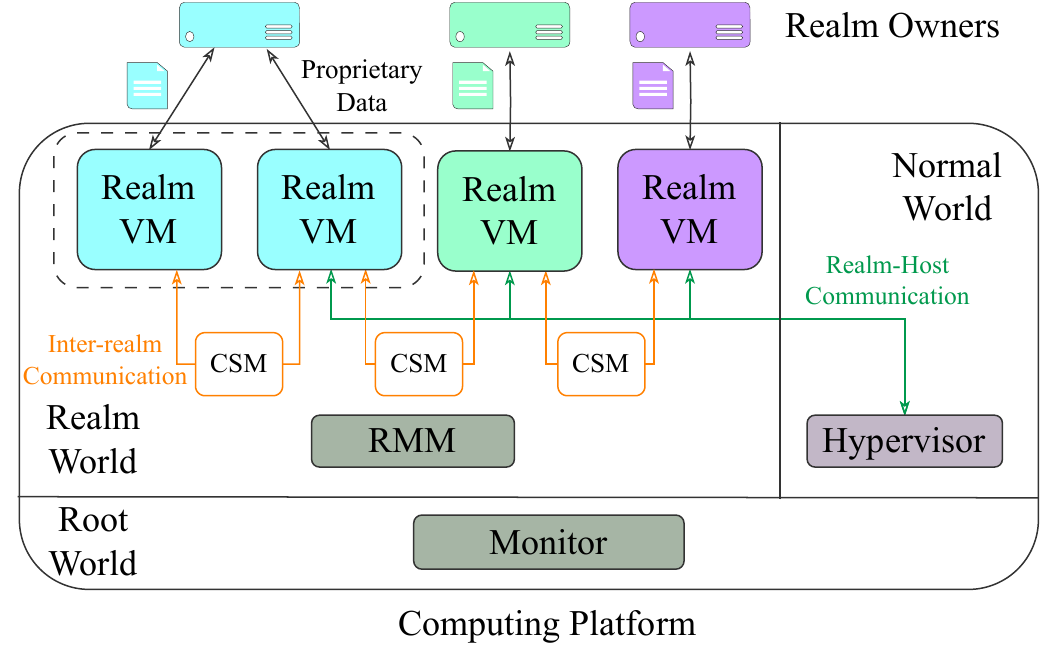}
   \caption{\name System Model}
    \label{fig:systemmodel}
\end{figure}
\names system model is illustrated in \cref{fig:systemmodel}.
\name is designed to operate across both edge platforms (\eg smartphones) and cloud platforms that support the Armv9-A architecture with CCA extensions.
The platform boots the CCA firmware, consisting of the RMM and the Monitor, along with a hypervisor responsible for managing system resources.
The hypervisor and all services running under its control are collectively referred to as the host.
Realm owners are entities external to the computing platform that provide confidential services, either locally (when the platform is an edge device) or remotely (in the cloud scenario).
Examples of such services include model inference~\cite{siby2024guarantee,zhang2024tsdp,moon2025asgard}, private LLM agents~\cite{wu2024isolategpt,abdollahi2025early}, digital rights management (DRM), and authentication services~\cite{vmcoreandroid}. 

Realm owners are mutually untrusted with respect to one another and to the host; each seeks to protect its proprietary data from all other entities.
Realm owners deploy realm(s) to deliver their services while preserving the confidentiality of their proprietary data from other entities.
A single realm owner may further compartmentalize its service into multiple realms for stronger isolation (Cyan realms in \cref{fig:systemmodel}).
Realm owners may agree to collaborate and share data only through \textit{attested realms and explicitly established CSM regions between those attested realms}.
The use of realms to deliver confidential services and host proprietary data of external entities (realm owners) adhere to the design principle of CCA~\cite{ccasite,armccasecurity} and align with other system models such as Android Virtualization Framework~\cite{avfarchitecture} and ASTER~\cite{kuhne2024aster}.
\name assumes that all entities in the system trust the device’s hardware and the CCA firmware. 
Finally, the hypervisor is assumed to provision the necessary resources (\eg memory and CPU time) to ensure forward progress of realms. 
%%%%%%%%%%%%%%%%%%%%%%%%%%%%%%%%%%%%%%%%%%%%%%%%%%%%%%%%%%%%%%%%%%%%%%
\subsection{Threat Model}\label{sec:adversarymodel}
Following the Arm CCA threat model~\cite{armccasecurity}, \name assumes that the host and all realms are mutually untrusted.
For any given realm, both the host and other realms may attempt to compromise the confidentiality or integrity of its protected memory or vCPU state.
This includes collaborating realms, which may try to read or modify memory outside the mutually agreed CSM boundaries.
A realm may also attempt to obtain additional resources (\eg granules) beyond those delegated by the host or retain resources longer than permitted, which may result in denial-of-service (DoS) attack against the host or other realms.
Introducing CSM support creates additional attack vectors.
Because \name exposes CSM creation and access to existing ones as a generic service available to all realms (including malicious realms), an adversarial realm might attempt to impersonate an attested peer to gain unauthorized access to an existing CSM.
Conversely, it may create a fake CSM region and trick other realms into using it, thereby compromising the confidentiality or integrity of shared data.
Such an adversarial realm may belong to a competing realm owner or be instantiated by the host itself with the intent of targeting a specific CSM configuration.
Physical attacks and microarchitectural side-channel attacks are out of scope.
The platform is assumed to support secure boot, ensuring the trusted loading and execution of all EL3 and EL2 components and preventing boot-time compromise.
Hardening of higher-level protocols and interfaces used by realms for inter-CVM communication is considered orthogonal to \names contribution.
%%%%%%%%%%%%%%%%%%%%%%%%%%%%%%%%%%%%%%%%%%%%%%%%%%%%%%%%%%%%%%%%%%%%%%
\iffalse
\subsection{Design Challenges}\label{sec:Design Challenges}
Design of \name faces two fundamental challenges.
\textbf{C1:} 
In CCA, the RMM is responsible for validation and isolation, while the hypervisor retains control over physical resource management. This division keeps the RMM small—and therefore easier to trust—but also complicates the introduction of CSM. Enabling CSM requires precise cooperation among the RMM, the hypervisor, and the participating realms, while preserving the semantics of realm creation, memory delegation, isolation, and teardown. CSM extensions must integrate cleanly across all layers of CCA without inflating the TCB. \name follows a minimalist design philosophy, reusing existing mechanisms for memory management, attestation, and isolation whenever possible and introducing only the essential new abstractions required to support CSM.
\textbf{C2:} Introducing CSM breaks the traditional disjoint memory model of CCA, thus \name must enforce that only authorized realms gain access to a CSM region while maintaining the isolation guarantees as CCA for all non-CSM memory.
This includes preventing impersonation, unauthorized sharing, and attempts by peer realms to access private memory outside the CSM bounds. Our in-depth security analysis in \cref{sec:securityanalysis} demonstrates how \name addresses these attack surfaces.
\fi
%%%%%%%%%%%%%%%%%%%%%%%%%%%%%%%%%%%%%%%%%%%%%%%%%%%%%%%%%%%%%%%%%%%%%%
\section{\name}\label{sec:Design}
In this section, we introduce \name. We begin by describing the setup of each realm in \cref{sec:realm setup}. We then present the full lifecycle of CSM in \cref{sec:csmlifecycle}, and finally detail several key design components, including the realm identifier (\cref{sec:realmidentifier}) and physical memory management (\cref{sec:physicalmemory}).
\subsection{Realm Setup}\label{sec:realm setup}
\begin{table*}[t]
  \centering
  \caption{New and modified commands introduced by \name}
  \resizebox*{\linewidth}{!}{%
    \begin{tabular}{lll}
      \toprule
      \textbf{Type} & \textbf{Name} & \textbf{Description} \\
      \midrule

      % ===========================================================
     \multicolumn{3}{l}{\textbf{\makebox[2.33cm][l]{Caller: Hypervisor} Callee: RMM}}  \\
      \cmidrule(lr){1-3}

      New RMI &
      \texttt{err = RMI\_APT\_CREATE(PA\textsubscript{RD},PA\textsubscript{APT})} &
      Allocates  the granule at physical address \texttt{PA\textsubscript{APT}} for the realm's APT.\\

      New RMI &
      \texttt{err = RMI\_APT\_DESTROY(PA\textsubscript{RD},PA\textsubscript{APT})} &
      Destroys the realm’s APT. This command is only invoked during realm destruction.\\
      Modified RMI &
          \texttt{err = RMI\_DATA\_CREATE\_UNKNOWN(PA\textsubscript{RD},PA\textsubscript{D},IPA\textsubscript{D})} &
      Maps a data granule \texttt{D} at physical address \texttt{PA\textsubscript{D}} into intermediate physical address \texttt{IPA\textsubscript{D}}.    \\
      & & \name ensures that if \texttt{IPA\textsubscript{D}} is in a CSM region, new mappings are created for other participating realms. \\
      Modified RMI &
      \texttt{err = RMI\_DATA\_DESTROY(PA\textsubscript{RD},IPA\textsubscript{D})} &
      Wipes and unmaps the data granule in \texttt{IPA\textsubscript{D}}.\\
      & & \name ensures that if \texttt{IPA\textsubscript{D}} is in a CSM region, it is unmapped for other participating realms. \\
      % ===========================================================
      \addlinespace[0.4em]
     \multicolumn{3}{l}{\textbf{\makebox[2.33cm][l]{Caller: P-realm} Callee: RMM}} \\
      \cmidrule(lr){1-3}
      New RSI &
      \texttt{ID\textsubscript{CSM} = RSI\_CSM\_CREATE(IPA\textsubscript{base-P},Size\textsubscript{CSM})} &
      Creates a CSM region beginning at \texttt{IPA\textsubscript{base-P}} with size \texttt{Size\textsubscript{CSM}}.  Returns CSM identifier \texttt{ID\textsubscript{CSM}}. \\

    New RSI &
      \texttt{ID\textsubscript{CSM-PC} = RSI\_CSM\_SHARE(ID\textsubscript{CSM},ID\textsubscript{C},P)} &
     Shares a CSM region \texttt{ID\textsubscript{CSM}} with a C-realm identified by \texttt{ID\textsubscript{C}} and access permission \texttt{P}. Returns the sharing identifier \texttt{ID\textsubscript{CSM-PC}}. \\

    New RSI &
      \texttt{err = RSI\_CSM\_REVOKE(ID\textsubscript{CSM-PC})} &
      Revokes access to an already existing sharing identified by \texttt{ID\textsubscript{CSM-PC}}. \\
      New RSI &
      \texttt{err = RSI\_CSM\_DESTROY(ID\textsubscript{CSM})} &
      Destroys a CSM region identified by \texttt{ID\textsubscript{CSM}}.\\
      % ===========================================================
      \addlinespace[0.4em]
       \multicolumn{3}{l}{\textbf{\makebox[2.33cm][l]{Caller: C-realm} Callee: RMM}} \\
      \cmidrule(lr){1-3}
      New RSI &
      \texttt{err = RSI\_CSM\_RESERVE(ID\textsubscript{CSM-PC},IPA\textsubscript{base-C},Size\textsubscript{CSM})} &
      Reserves a region begins at \texttt{IPA\textsubscript{base-C}} with size \texttt{Size\textsubscript{CSM}} for sharing identified by \texttt{ID\textsubscript{CSM-PC}}.\\
      New RSI &
      \texttt{err = RSI\_CSM\_ATTACH(ID\textsubscript{CSM-PC})} &
      Maps the CSM with sharing identifier \texttt{ID\textsubscript{CSM-PC}} into the corresponding reserved region,  delegating access to the CSM.  \\

      New RSI &
      \texttt{err = RSI\_CSM\_DETACH\_AND\_FREE(ID\textsubscript{CSM-PC})} &
      Unmaps the CSM with sharing identifier \texttt{ID\textsubscript{CSM-PC}} from the C-realm private address space and frees the previously reserved region.\\
      % ===========================================================
      \addlinespace[0.4em]
       \multicolumn{3}{l}{\textbf{\makebox[2.33cm][l]{Caller: RMM} Callee: Hypervisor}} \\
      \cmidrule(lr){1-3}
      New Exit Reason &
      \texttt{REC\_EXIT\_C\_REALM\_CSM(IPA\textsubscript{base-C},Size\textsubscript{CSM})} &
      Notifies the hypervisor about a new CSM region in the C-realm’s protected address space.\\
      New Exit Reason &
      \texttt{REC\_EXIT\_P\_REALM\_CSM(IPA\textsubscript{base-P},Size\textsubscript{CSM})} &
      Notifies the hypervisor about a new CSM region in the P-realm’s protected address space.\\
       New Exit Reason &
      \texttt{REC\_EXIT\_REALM\_REMOVE\_CSM(IPA\textsubscript{base-P},Size\textsubscript{CSM})} &
      Notifies the hypervisor to remove a CSM region from realm's protected address space.\\

      \bottomrule
    \end{tabular}
  }
  \caption*{\footnotesize The physical address of RD \texttt{PA\textsubscript{RD}} must be provided within each RMI command.}
  \label{tab:RMI_RSI}
\end{table*}
\begin{figure*}[t]
    \centering
    \includegraphics[width=1\linewidth]{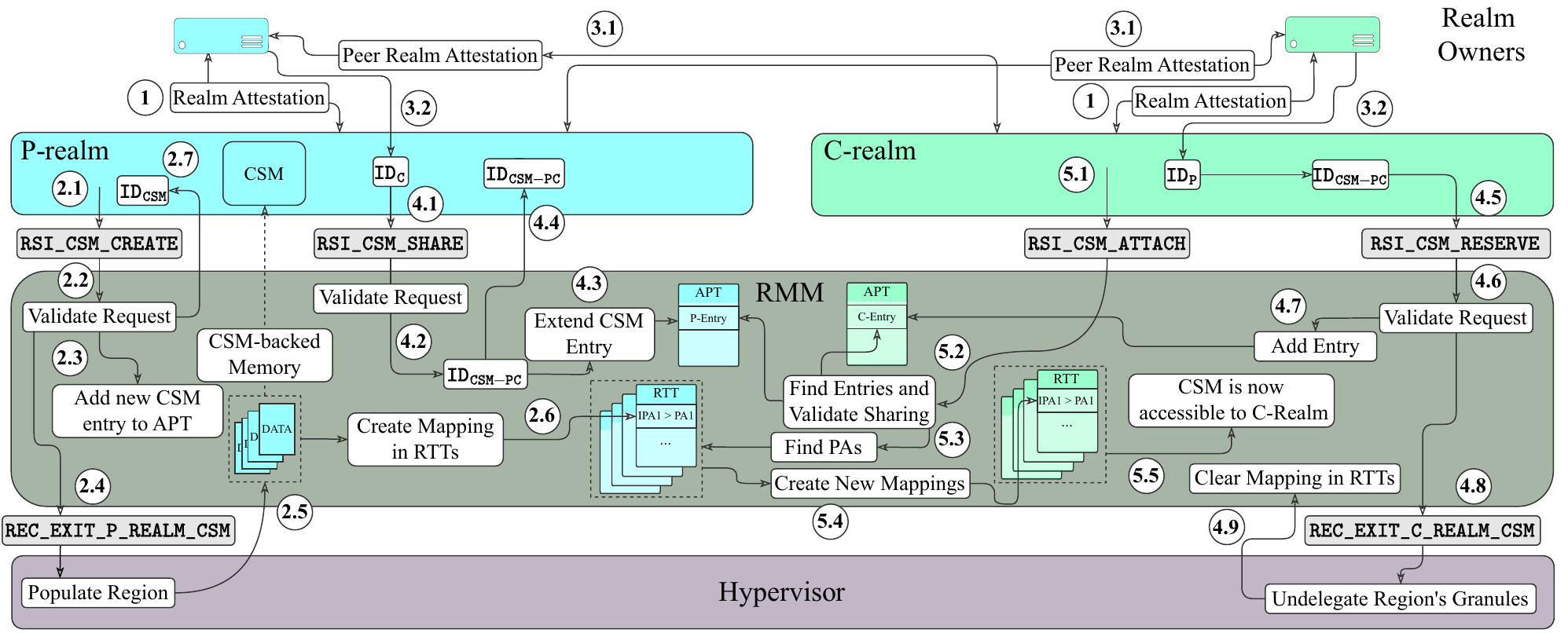}
   \caption{Overview of \name. 
   }
    \label{fig:systemoverview}
\end{figure*}
%%%%%%%%%%%%%%%%%%%%%%%%%%%%%%%%%

At a high level, \name extends Arm CCA with support for CSM, enabling realms to securely create, manage, and attach to CSM regions. The RMM serves as the trusted core of \name: it implements validation, enforces access control, and orchestrates all CSM-related operations.  \name introduces new RSI commands to the RMM, which exposes the CSM-related services to realms (see \cref{tab:RMI_RSI}).

\para{Ownership Model} 
In \name each CSM region has exactly one creator and lifetime manager, referred to as its Provider realm (P-realm). The P-realm (1) contributes a portion of its protected address space to instantiate the CSM region, (2) grants or revokes other realms’ access to it, and (3) determines the access permissions types (\eg read-only or read-write) of each participating realm. Realms that later join an existing CSM region are referred to as Consumer realms (C-realms).
Crucially, these roles are per-region and not global: a realm may act as a P-realm for some CSM regions and a C-realm for others.

\para{Allocation Semantics} \name preserves the hypervisor’s control over physical memory management for both CSM and non-CSM regions, aligning with Arm CCA’s design principles. 
 Since a CSM region is shared across multiple realms, memory-management flows must diverge from those used for regular private memory of realms. Accordingly, the RMM notifies the hypervisor whenever CSM regions are created, shared, or removed (through vCPU exits, see \cref{tab:RMI_RSI}). The hypervisor uses this information to populate the P-realm’s CSM region while ensuring that C-realms do not receive additional physical granules beyond what the P-realm explicitly shares (see \cref{sec:physicalmemory} for further details).  \name, however, guarantees that once a CSM region is established, all participating realms observe an identical and consistent view of the underlying physical granules. 

 \para{Access Policy Table}
%The RMM in \name is responsible for managing CSM regions and enforcing access control. 
The RMM in \name requires to effectively track the state and ownership of each CSM region. In particular, whenever a realm issues a CSM-related RSI command, the RMM must consult—and potentially update—the corresponding metadata to ensure that all security checks are correctly enforced. In CCA, however, the RMM does not store per-realm metadata directly in its internal memory; instead, it relies on metadata granules that the hypervisor delegates to each realm (\eg RTT granules).
Following this design principle, \name introduces a new metadata structure, Access Policy Table (APT), associated with each realm.  The APT allows the RMM to record metadata for all CSM regions mapped into a realm’s protected address space. APT record one entry per CSM region, which is either  P-realm entry or C-realm entry (depending on the role of the realm in that CSM). C-realm entries store information such as base address, size, the identifier of the owner (P-realm), while P-realm entries keep additional metadata such as permission type of C-realms authorized to access that CSM. To prevent race conditions or deadlocks, \name reuses the RMM’s existing locking mechanisms~\cite{rmmlock}, ensuring that multiple vCPUs cannot access or modify realm’s APT concurrently.
%
%For each realm, the hypervisor delegates a granule to serve as its APT using the \texttt{RMI\_APT\allowbreak\_CREATE} command, and the APT is reclaimed during realm teardown using \texttt{RMI\_APT\allowbreak\_DESTROY} (\cref{tab:RMI_RSI}). Although the hypervisor allocates and delegates the APT granule, it cannot affect its contents. APT entries can only be created, updated, or removed in response to RSI commands.
%

\para{Realm Initialization} 
\name adapts the realm initialization flow of CCA, in which the hypervisor issues RMI commands \cite{rmmspec} to populate the initial realm’s image and delegate realm’s metadata granules. The metadata granule introduced by \name (\ie APT) is similarly created and initialized through a new RMI command, \texttt{RMI\allowbreak\_APT\allowbreak\_CREATE} (see \cref{tab:RMI_RSI}). At this stage, the APT is allocated for each realm within the RMM’s internal structures but does not yet contain any configuration.
Once initialization is complete, the hypervisor begins scheduling the vCPUs of each realm. At this point,  realms can communicate with external entities, including their respective realm owners. The realms establish TLS channel and prove  them self to their realm owner via attestation (step \circled{1} in \cref{fig:systemoverview}).
\subsection{CSM Lifecycle}\label{sec:csmlifecycle}
\para{Creation}
A realm can create arbitrary CSM regions in its private address space via \texttt{RSI\_CSM\allowbreak\_CREATE}, providing the region’s start address and size (see \cref{tab:RMI_RSI}). While \name does not restrict the number of CSM regions a realm may create, each region must (1) be granule-aligned and (2) not overlap with any existing CSM region in the realm’s private address space. 
%Granule alignment is required to maintain granule-level sharing semantics, and non-overlap ensures that a C-realm cannot transfer or extend sharing of the to other realms, preserving the ownership model described earlier.
If these conditions are satisfied, the RMM registers this region within the APT of P-realm, notifies the hypervisor about the new CSM region via vCPU exit \texttt{REC\_EXIT\_P\allowbreak\_REALM\allowbreak\_CSM}. The hypervisor then populate the entire CSM range within the P-realm's private address space (see \cref{sec:physicalmemory}). For every new delegated granule, the RMM creates the corresponding entry in the P-realm's RTTs. The RMM later returns a unique region identifier \texttt{ID\textsubscript{CSM}} to the realm (steps \circled{2.1} to \circled{2.7} in \cref{fig:systemoverview}).

\para{Pre-Sharing}
Before sharing CSM, realms must be able to securely identify their peers. To enable this, \name adapts the attestation primitives of CCA and delegates peer realm attestation to the realm owners. Each realm owner attests the peer realm to verify that the expected software stack is running within it (step~\circled{3.1}).
In \name, the RMM assigns each realm a unique identifier that is used to refer to that realm during subsequent CSM-related operations. The RMM also reports these identifiers as a separate claim in the attestation token. As a result, realm identifiers can be reliably recovered by realm owners during peer realm attestation. After validating the attestation report (\ie checking that realm's claims match the expected values), realm owners distribute the validated peer realm's identifier to their respective realms (step \circled{3.2}). From this point onward, both realms know each others identifiers.
In \cref{sec:realmidentifier}, we provide a detailed explanation on realm identifier and the way \name builds it.

\para{Sharing} To allow sharing CSM between realms,
\name adopts a simple but strong rule: a realm can attach to an existing CSM region only if both the P-realm and the C-realm explicitly agree to share and attach, respectively. 
%
%Assuming that both realms know each other’s unique identifiers and the intended CSM size in advance; 
P-realm can initiate sharing by issuing \texttt{RSI\_CSM\allowbreak\_SHARE}, providing the region identifier \texttt{ID\textsubscript{CSM}}, the desired access permissions \texttt{P}, and the identifier of the  C-realm \texttt{ID\textsubscript{C}}. The RMM validates these inputs and, upon success, records a sharing identifier in the P-realm’s APT entry associated to the CSM region, marking that the P-realm has agreed to share the CSM with the specified C-realm. It returns the sharing identifier \texttt{ID\textsubscript{CSM-PC}} to the P-realm (steps \circled{4.1} to \circled{4.4}). The identifier is a deterministic concatenation of the P-realm and C-realm identifiers, combined with a counter to distinguish multiple shared regions between the same pair of realms. 

On the C-realm side, it can regenerate the sharing identifier due to its deterministic format. The C-realm must then reserve a region of identical size in its own private address space using \texttt{RSI\allowbreak\_CSM\allowbreak\_RESERVE}. It provides the region’s start address, size, and the region identifier \texttt{ID\textsubscript{CSM}} to the RMM. The RMM verifies that the reserved range (1) is granule-aligned and (2) does not overlap with any existing CSM region already present in the C-realm’s address space.
If the checks succeed, the RMM adds a corresponding entry to the C-realm’s APT, indicating that the C-realm has agreed to attach to the region shared by the designated P-realm. The RMM then notifies the hypervisor via \texttt{REC\_EXIT\_C\allowbreak\_REALM\allowbreak\_CSM}, prompting the hypervisor to undelegate any previously assigned granules in the reserved CSM range of the C-realm (steps \circled{4.5} to \circled{4.9}).

\para{Access} To gain access to the shared region, the C-realm must finally issue \texttt{RSI\allowbreak\_CSM\allowbreak\_ATTACH} with the sharing identifier \texttt{ID\textsubscript{CSM-PC}} provided. The RMM first locates the corresponding entries in both the P-realm’s and the C-realm’s APTs; the presence of both entries confirms that the P-realm has previously agreed to share the region and the C-realm has agreed to attach. 
\name does not require the CSM region to appear at the same base IPA address in each realm’s private address space. However, both realms must use an identical size for the region; this ensures that neither realm gains unauthorized access to memory outside the mutually agreed-upon CSM boundaries.
If these conditions are satisfied, the RMM begins constructing the required mappings in the C-realm’s address space. 
%Assuming the hypervisor has already backed the CSM with sufficient physical resources (\cref{sec:physicalmemory}), 
To achieve that, the RMM  walks through the P-realm’s RTTs to locate the physical addresses corresponding to the CSM’s range and creates equivalent mappings in the C-realm’s RTTs (steps \circled{5.1} to \circled{5.5}). During this process, the RMM applies the access permissions previously specified by the P-realm to all new RTT entries of C-realm. Once these mappings are established, the C-realm can access the CSM region according to the assigned permissions.
%From this point onward, the RMM syncs the hypervisor memory delegation and undelegates for both realms. For example, if hypervisor 
% 

\para{Revocation \& Destruction} The P-realm retains its control over sharing of the CSM with the C-realm. It can  issue \texttt{RSI\allowbreak\_CSM\allowbreak\_REVOKE}, which removes the mapping from the C-realm's RTT, and clear the C-realm's identifier from the associate APT entry in P-realm APT. A P-realm can also destroy a CSM region via \texttt{RSI\allowbreak\_CSM\allowbreak\_DESTROY} with the region identifier provided. In return the RMM, repeats the handler of \texttt{RSI\allowbreak\_CSM\allowbreak\_REVOKE} command for every peer C-realm and finally  destroys the associated APT entry. The RMM then notifies the hypervisor about the destruction of CSM sharing in the P-realm's private address space via \texttt{REC\_EXIT\allowbreak\_REMOVE\allowbreak\_CSM}.
The C-realm can also unmap and free its address space from the CSM via \texttt{RSI\_DETACH\allowbreak\_AND\allowbreak\_FREE} with the sharing identifier provided. In return, the RMM remove mapping from C-realm's RTT, removes the associated entry in the APT of C-realm, and notifies the hypervisor about the destruction of CSM sharing in the C-realm's private address space via \texttt{REC\_EXIT\allowbreak\_REMOVE\allowbreak\_CSM}. 
%

%\para{Access Permissions}
%In \name, the P-realm can assign access permissions—such as read-only or non-executable—to the CSM. The P-realm can share its data with higher confidence about its integrity protection against both intentional and unintentional modifications by other realms. Enforcing restrictive permissions can also be used in the design of the interface between peer realms, hardening against adversaries, a principle shown effective in prior works~\cite{yu2022elasticlave,kuhne2024aster}. 
%%%%%%%%%%%%%%%%%%%%%%%%%%%%%%%%%%%%%%%%%%%%%%%%%%%%%%%%%%%%%%%%%%%%%%
\subsection{Realm Identifier}\label{sec:realmidentifier}
In the traditional model of CCA , all RMM services (exposed as RSI and RMI commands) are \textit{local} to a single realm. By contrast, \name extends the RMM to support services \textit{between} realms, enabling realms to share CSM with one another. This fundamentally requires realms to identify and refer to each other securely. For example, a realm invoking \texttt{RSI\_CSM\allowbreak\_SHARE} must specify the identifier of the target C-realm. The RMM must be able to reliably interpret these identifiers to enforce access control throughout the CSM lifecycle. If realm identifiers were forgeable or manipulable (\eg by a malicious hypervisor), the security of
CAEC would be fundamentally compromised. Thus, \name requires
a secure, unforgeable mechanism for realm identification.

\para{CCA Identifier Model} The RMM design  in CCA provides no such system-wide realm identifier, as the RMM was not designed to support inter-realm operations. The RMM in CCA distinguishes realms solely by the physical address of their RD, which the hypervisor supplies with each RMI invocation. 
However, this identifier is unsuitable to be used as identifier between realms.
First, exposing RD physical addresses as the identifier to entities such as realm owners and their associated realms meaning that they obtain information about the physical memory layout of the system, which violates virtualization principles.
Second, since the hypervisor can control physical memory layout, it provides no uniqueness guarantee: a malicious hypervisor could terminate the current realm and instantiate a malicious one at the same RD address, causing both to appear to have the same identifier. This enables a classic time-of-check-to-time-of-use (TOCTOU) attack in which the hypervisor can replace a legitimate realm with a malicious one while preserving the same identifier.
In summary, none of the realm-related arguments available in the current CCA design satisfies both essential properties: (1) not leaking system-level information, and (2) being protected from hypervisor manipulation. Candidates such as the RD physical address, the Realm Personalization Value (RPV), and the Virtual Machine ID (VMID)~\cite{rmmspec} all fail to meet both criteria.

\para{Realm Identifier in \name} The RMM in \name assigns a unique system-wide identifier to each newly created realm and maintains a registry of these identifiers. These identifiers reveal no information about the underlying system and are safe from hypervisor manipulation, as each realm receives a fresh identifier upon creation. The RMM further reports realm identifier as a separate claim within each attestation token, creating a cryptographic binding between the identifier and other claims—such as the RIM—within the attestation token. As a result, the realm identifier becomes an attestable property that third parties can verify alongside other claims. The attestation token in \name can prove (1) the realm’s software and platform configuration (as in CCA), and (2) the authenticity of the realm identifier (new in \name). Consequently, realm owners can use attestation to obtain authenticated identifiers of peer realms and safely distribute them to their own realms for subsequent CSM-related operations.

%further extends the CCA attestation token to include the realm identifier as a separate claim.

%%%%%%%%%%%%%%%%%%%%%%%%%%%%%%%%%%%%%%%%%%%%%%%%%%%%%%%%%%%%%%%%%
\subsection{Physical Memory Allocation}\label{sec:physicalmemory}
When a new CSM region is registered, the RMM notifies the hypervisor through the vCPU exits \texttt{REC\allowbreak\_EXIT\_P\_REALM\allowbreak\_CSM} and \texttt{REC\allowbreak\_EXIT\allowbreak\_C\allowbreak\_REALM\allowbreak\_CSM}. These notifications allow the hypervisor to finalize the physical memory configuration required for subsequent system operations. If the request corresponds to the creation of a CSM region by the P-realm, the hypervisor delegates granules for the entire CSM range, if they are not already delegated, ensuring that the region is fully populated and accessible to the P-realm. Conversely, if the request corresponds to attaching to an existing CSM region by a C-realm, the hypervisor undelegates any previously delegated granules that fall within the CSM range of that realm's protected address space. This mechanism guarantees that the CSM region is instantiated exactly once in physical memory, maintaining consistency across participating realms and preventing redundant allocation.

For CSM creation, the hypervisor performs the following steps for each granule in the CSM range. It first issues \texttt{RMI\allowbreak\_RTT\_READ\allowbreak\_ENTRY} to check whether the target IPA is already populated. If it is not populated, the hypervisor issues \texttt{RMI\_GRANULE\_DELEGATE} followed by \texttt{RMI\allowbreak\_DATA\allowbreak\_CREATE\allowbreak\_UNKNOWN} to delegate a physical granule and create the corresponding mapping in the realm's RTT, respectively. 
For CSM attachment, the hypervisor again invokes \texttt{RMI\_RTT\allowbreak\_READ\allowbreak\_ENTRY}. If the IPA is already populated, it reclaims the granule using \texttt{RMI\allowbreak\_DATA\allowbreak\_DESTROY} and \texttt{RMI\_GRANULE\allowbreak\_UNDELEGATE}.
The hypervisor also ensures that the RTTs of both the P-realm and the C-realm exist for the CSM range. If necessary, it creates them by issuing \texttt{RMI\allowbreak\_GRANULE\allowbreak\_DELEGATE} followed by \texttt{RMI\allowbreak\_RTT\_CREATE}.
As outlined in the adversary model (\cref{sec:adversarymodel}), we assume that the hypervisor correctly allocates all required resources to ensure uninterrupted realm execution. However, failure to allocate these granules does not compromise the security guarantees of \name.
%%%%%%%%%%%%%%%%%%%%%%%%%%%%%%%%%%%%%%%%%%%%%%%%%%%%%%%%%%%%%%%%%%%%%%
\section{Security Analysis}\label{sec:securityanalysis}
In this section we provide an in-depth analysis of different attacks vectors introduced in threat model (\cref{sec:adversarymodel}), explaining how \name remains protected against these threats.

\para{CSM Protection}
As discussed in \cref{sec:backgroundarmcca}, the hardware-enforced GPC mechanism ensures that any access to realm-world memory from either the normal world or the secure world is strictly prohibited.
Since the RMM enforces that all CSM regions reside the realm world, CSM regions are inherently protected from direct access attempts originating from normal world and secure world actors.

\name defends the CSM  against malicious realms through two complementary mechanisms:
~(1) A system-wide, attestation-integrated realm identifier, and
~(2) RMM-enforced access control checks within all CSM-related RSI commands.
First, reporting realm's identifier as a separate claim in the attestation token creates a cryptographic binding between realm identifier and its other claims such as RIM. Realm owners verify these claims and proceed only if they match the expected configuration, after which they provision the validated identifier to their associated realm. A malicious realm, whether instantiated by a competing realm owner or by the hypervisor, cannot bypass this attestation check because its software content (and thus its RIM) will not match the expected values. Consequently, such a realm cannot impersonate an attested peer, attach to an existing CSM, or create a fraudulent CSM region to lure honest realms.
Second, after attestation, all CSM participation requests are subject to explicit access-control checks performed by the RMM. Both the P-realm and the C-realm must independently reference each other using their attested identifiers when issuing CSM-related RSI commands. This bidirectional referring rule ensures that, although a malicious realm may arbitrarily invoke CSM-related RSI commands, it cannot establish a CSM with a realm to gain access to a portion of its address space, unless the peer realm explicitly agrees.

Finally, the RMM performs required cache and TLB maintenance whenever mappings change. These operations flush stale translations, ensuring that no realm can retain access to the CSM after its permissions have been revoked. 

\para{Non-CSM Protection}
\name guarantees that neither the P-realm nor the C-realm can access each other’s protected memory outside the agreed-upon CSM region.
During the CSM sharing phase, the RMM validates that sharing occurs only at granule-aligned boundaries and for the exact size confirmed by both realms, thereby preventing malicious  access outside the CSM.  

\para{Ownership Protection}
When a C-realm issues \texttt{RSI\_CSM\allowbreak\_RESERVE}, the RMM explicitly verifies that the requested region does not overlap with any existing mappings. This restriction prevents a C-realm from delegating access to a CSM region to another realm without the P-realm’s consent. By ensuring that all sharing relationships originate from the P-realm, ownership and access remain fully traceable. Such control is crucial for allowing the P-realm to reason about which realms have visibility into the shared memory and for guaranteeing that, once access is revoked, no realm can retain or reacquire access to the CSM.

\para{Host Protection}
The RMM in \name imposes no constraints on how the hypervisor manages CPU scheduling or memory allocation for either CSM or non-CSM regions. As a result, realms cannot exploit \names services to obtain more memory or vCPU resources than those explicitly provisioned by the hypervisor. This design protects the hosting system from DoS attacks originating from malicious or misbehaving realms.
%%%%%%%%%%%%%%%%%%%%%%%%%%%%%%%%%%%%%%%%%%%%%%%%%%%%%%%%%%%%%%%%%%%%%%
\section{Implementation}\label{Implementation}
\begin{table}[t]
  \centering
  \caption{Line of code added to components in \name}
  \label{tab:TCB}
    \begin{tabular}{lc}
      \hline
      \textbf{TCB Component} & \textbf{Lines of Code (Extension)} \\
      \hline
      RMM (v0.5.0)\cite{RMM}     & 1825 (6\%)   \\

      \hline
      \textbf{Non-TCB Component} & \textbf{Lines of Code}  \\
      \hline
      kvmtool-cca (v3/cca)\cite{kvmtool-cca}   & 247 \\
      Linux KVM (v5+v7)\cite{linux-cca}        & 525 \\
      CSM Driver                              & 594 \\
      \hline
    \end{tabular}%
\end{table} 
We implement and evaluate \name with both functional and performance prototypes of CCA. 
Our implementation includes  both host-side and realm-side extensions to support the creation and use of CSM regions. 

\para{Software Stack}
We use unmodified Trusted Firmware-A \cite{TF-A} (v2.11) as the Monitor. We adapt the Trusted Firmware reference implementation of the RMM \cite{RMM} (v0.5.0), extending it to support the new and modified commands introduced by \name (\cref{tab:RMI_RSI}). On the host side, we base our hypervisor on linux-cca \cite{linux-cca} (v5+v7) and the virtual machine manager on kvmtool-cca \cite{kvmtool-cca} (v3/cca).
We extend the Linux KVM (Kernel-based Virtual Machine) module~\cite{kvm,dall2014kvm} to support the new RMI commands and to delegate physical memory to CSM regions in response to the newly defined vCPU exits from the RMM (\cref{tab:RMI_RSI}). We modify kvmtool to insert a node into the guest device tree,  reserving a portion of its private address space to be used as the CSM.
On the guest side, we add CSM driver to linux-cca \cite{linux-cca} (v5+v7). This driver discovers the reserved region from the device tree, issues the appropriate RSI commands to create or attach to CSM regions, and exposes CSM to user space as a character device. The same driver is used by both P-realms and C-realms.

\Cref{tab:TCB} summarizes the code changes introduced by \name, measured in lines of code (LoC). \name adds 1,825 LoC to the RMM (29k LoC), resulting in approximately a 6\% increase in its size. The RMM version used in our prototype (v0.5.0) targets CCA v1.0, which lacks planned features such as device assignment and planes (\cref{sec:futureextension}). These forthcoming features are expected to significantly increase the RMM’s size, making \name’s relative contribution to the TCB even smaller over time.
On the host side, \name contributes 247 LoC to kvmtool, of which 213 LoC implement the PCI device used in our experimental setup (\cref{sec:communicationbechmark}), and 34 LoC correspond to core \name functionality. \name also adds 525 LoC to the Linux KVM. On the guest side, the CSM driver adds 594 LoC.

\para{Functional Prototype}
At the time of writing, two functional prototypes of Arm CCA are publicly available.
Arm’s Fixed Virtual Platform (FVP) and Linaro’s QEMU~\cite{QEMUlinaro} both provide emulated CCA-compatible hardware.
We adopt FVP as our functional prototype because it is Arm’s official release, fully aligned with the CCA specification~\cite{armccaenablement,fvp}, and widely used in prior work~\cite{wang2024cage,zhang2023shelter,sridhara2024acai,siby2024guarantee,abdollahi2025early}.
FVP models key hardware components of an Arm system, including the processors, cache hierarchy, bus traffic, and memory subsystem.
It is instruction-accurate, meaning that it correctly executes architecturally valid code, but it does not model cycle-accurate timing or the performance characteristics of real processors~\cite{FastModel,fvp}.
Consequently, FVP is well-suited for evaluating the feasibility and functional correctness of \name on Arm CCA hardware, but it cannot be used for timing or performance measurements.

\para{Performance Prototype}
%To prototype Arm CCA in the absence of real hardware, prior works have adopted a similar strategy of emulating a dual-world environment using existing boards.
At the time of writing, no commercial hardware implements the Arm CCA extensions. For performance prototyping, we therefore adopt OpenCCA~\cite{bertschi2025opencca}, the first open-source performance prototype of Arm CCA. Our performance evaluation can be easily reproduced as OpenCCA runs on a cost-effective hardware (\ie Radxa Rock 5B~\cite{ROCK5B}). 
OPENCCA constructs the realm world as a separate execution environment within the normal world. Because current off-the-shelf hardware lacks essential CCA extensions (\eg GPC), it cannot provide hardware-enforced isolated realm world. Nevertheless, performance prototyping remains feasible by emulating the realm world in software. This is achieved by splitting the normal-world software stack into two execution domains: one running the standard normal-world environment, and the other hosting the RMM and realm components. A modified Monitor then manages boot and context switching between these domains, effectively emulating transitions between the normal world and the realm world.

We acknowledge that such a prototype cannot fully capture the performance characteristics of real CCA hardware.
For example, the absence of GPC in the system registers and  memory system, can affect memory access behavior and caching, introducing discrepancies compared to a genuine CCA environment.
While we do not claim that our performance results precisely reflect the cost of \name on production-grade CCA hardware, this setup provides a best-effort approximation until such hardware becomes available.

%%%%%%%%%%%%%%%%%%%%%%%%%%%%%%%%%%%%%%%%%%%%%%%%%%%%%%%%%%%%%%%%%%%%%%
\section{Evaluation}\label{Evaluation}
In this section, we evaluate \name by addressing the following key questions:
\begin{itemize}
\item \textbf{Q1}: Is \name fully compatible with CCA hardware?
\item \textbf{Q2}: What are the performance gains introduced by \name?
\item \textbf{Q3}: What are the performance overheads introduced by \name?
\end{itemize}
%\item \textbf{Q2}: What is the performance overhead introduced by \name?

We answer \textbf{Q1} in \cref{sec:compatibility} 
%\textbf{Q2} in \cref{sec:overheadbenchmark}, 
, \textbf{Q2} in \cref{sec:communicationbechmark} and \cref{sec:datasharing}, and finally  \textbf{Q3} in \cref{sec:runtimecost} and \cref{sec:memorybench}.
%%%%%%%%%%%%%%%%%%%%%%%%%%%%%%%%%%%%%%%%%%%%%%%%%%%%%%%%%%%%%%%%%%%%%%
\subsection{Experimental Setting}
For functional prototype we set FVP to have two clusters, each with four cores supporting Armv9.2-A and 4GB of RAM. 
 All performance experiments are conducted on a Radxa Rock 5B+ board~\cite{ROCK5Bplus} equipped with 16GB of RAM and an 8-core processor (4× ARM Cortex-A76 and 4× ARM Cortex-A55).
   We always create realm VMs with one vCPU and pin each vCPU to a specific core, with highest scheduling priority given to the realm's vCPU process. For LLM-inference experiments, we use llama.cpp \cite{llamacpp} as our inference engine. 
%%%%%%%%%%%%%%%%%%%%%%%%%%%%%%%%%%%%%%%%%%%%%%%%%%%%%%%%%%%%%%%%%%%%%
\subsection{Compatibility with CCA Hardware}\label{sec:compatibility}
To evaluate the compatibility of \name with real CCA hardware, we repeated the data-sharing benchmark (\cref{sec:datasharing}) on FVP. We observed no runtime errors or stalls on FVP cores during CSM creation, inference execution, and CSM termination, demonstrating that \name can operate on Arm CCA–enabled hardware. In all experiments, we use the CSM driver to expose the CSM region to user-space processes.
%%%%%%%%%%%%%%%%%%%%%%%%%%%%%%%%%%%%%%%%%%%%%%%%%%%%%%%%%%%%%%%%%%%%%%

\subsection{Communication Benchmark}\label{sec:communicationbechmark}
\begin{figure*}[t]
    \centering

    % --- The Figure ---
    \includegraphics[width=\linewidth]{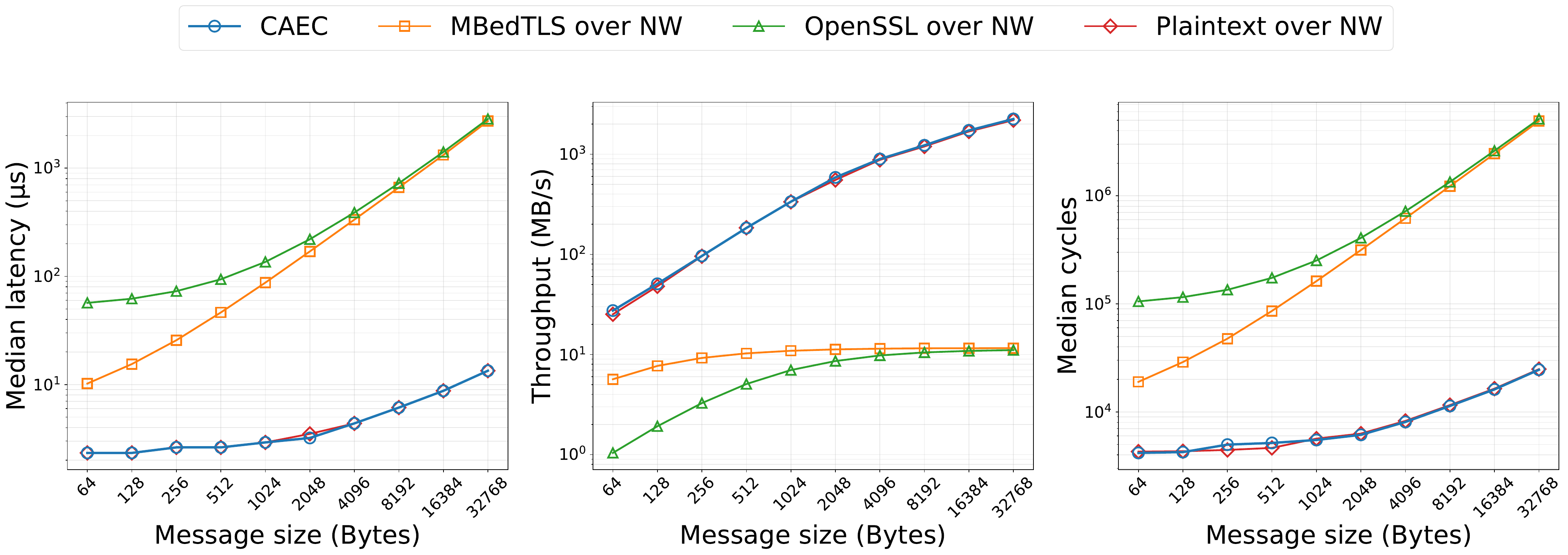}

    \vspace{0.5em}
    \begin{minipage}{0.31\linewidth}
        \centering  (a) Median latency for different message sizes.
    \end{minipage}
    \begin{minipage}{0.31\linewidth}
        \centering  (b) Communication throughput for different message sizes.
    \end{minipage}
    \begin{minipage}{0.31\linewidth}
        \centering (c) CPU usage per message size.
    \end{minipage}

    % --- Caption ---
    \caption{
        Communication cost between two realms across four modes:
        plaintext over NW shared memory, encrypted (OpenSSL/MbedTLS)
        with confidentiality and integrity, and \name.
    }
    \label{fig:all}
\end{figure*}
In this section, we evaluate the effectiveness of \name for communication between realms.
We assume two user-space programs running in separate realms exchange messages via shared memory.
The shared memory is provided either by \name or through a shared region in normal world.
In each experiment, one side writes  messages into the shared memory, and the other side reads them.
For the NW shared memory case, we evaluate two configurations:
(1) Encrypted communication, in which two sides employ an encryption/decryption protocol (using mbedTLS or OpenSSL) to each exchanged message; and
(2) Plaintext communication, in which no encryption is applied.
The encryption-based configuration provides confidentiality and integrity guarantees against an active NW adversary capable of intercepting or modifying shared-memory content.
In contrast, when using \name, encryption is unnecessary because the CSM region is inherently protected from both the NW and other realms.

\para{Results}
\Cref{fig:all} presents the results for CPU usage, latency (message delivery time), and throughput (data transferred per unit time). Across all metrics, \name consistently outperforms both encrypted NW-based configurations. Compared to OpenSSL, \name achieves 24$\times$–212$\times$ lower latency, 25$\times$–209$\times$ fewer CPU cycles, and 26$\times$–204$\times$ higher throughput. Against MbedTLS, \name delivers 4.4$\times$–203$\times$ lower latency, 4.5$\times$–200$\times$ fewer cycles, and 4.8$\times$–194$\times$ higher throughput. In all cases, the performance gap widens as message size increases.
These results highlight the \textit{substantial gain} achieved by the use of \name instead of NW shared memory for inter-realm communication. Because \name provides system-level protection for the CSM, it eliminates the need for encryption and its associated per-message costs.  \Cref{fig:all} also reports the performance of plaintext communication over NW shared memory. As the results indicate, \name achieves performance equivalent to plaintext NW communication, demonstrating that the previously observed differences between \name and encrypted modes over NW memory stem entirely from cryptographic processing rather than from differences in memory-access latency, caching effects, or other architectural factors.

For completeness, we note that in the encryption-based modes, each message consists of a small header—containing session\_id and seq—followed by the main payload.
The receiver verifies the header for integrity, checks that seq matches the expected value (ensuring ordering and replay protection). 
Each side allocates a buffer on the shared memory to acknowledge the latest sent/received massage. Synchronization between the two side is achieved through polling over these buffers.
Each experiment is repeated 1000 times to compute median values for latency and CPU usage, while the throughput measurement transfers 1000 messages sequentially. 
Finally, note that as kvmtool does not natively support sharing NW memory pages between two VMs, we extended it by implementing a new PCI device, with a design similar to ivshmem~\cite{ivshmem} in QEMU.
%%%%%%%%%%%%%%%%%%%%%%%%%%%%%%%%%%%%%%%%%%%%%%%%%%%%%%%%%%%%%%%%%%%%%%
\subsection{Data Sharing Benchmark}\label{sec:datasharing}
\begin{table*}[t]
  \centering
  \caption{Memory footprint comparison of inference services deployed in multiple realms.
  In the baseline, each realm hosts its own LLM; with \name, multiple realms share a single model.}
  \resizebox{\linewidth}{!}{%
  \begin{tabular}{cc|cc|cc}
    \toprule
    \multirow{4}{*}{\textbf{Model}} &
    \multirow{4}{*}{\textbf{Model Size (MB)}} &
    \multicolumn{2}{c|}{\textbf{2 Realms}} &
    \multicolumn{2}{c}{\textbf{3 Realms}} \\
    \cmidrule(lr){3-4}\cmidrule(lr){5-6}
    &  &
    \textbf{Memory Footprint (MB)} & \multirow{3}{*}{\textbf{Reduction (\%)}} &
    \textbf{Memory Footprint (MB)} & \multirow{3}{*}{\textbf{Reduction (\%)}} \\
    &  & \textbf{Baseline / \name} & & \textbf{Baseline / \name} & \\
     & & \textbf{Total (P-realm, C-realm)} & & \textbf{Total (P-realm, 2*C-realm)} & \\
    \midrule
    GPT-2-Q8 \cite{gpt2} & 177 &
    960 (480, 480) / 800 (490, 310) & 16.6\% &
    1440 (480, 960) / 1110 (490, 620) & 22.9\% \\
    GPT-2-Medium-Q8  \cite{gpt2-medium} & 437 &
    2000 (1000, 1000) / 1580 (1010, 570) & 21.0\% &
    3000 (1000, 2000) / 2150 (1010, 1140) & 28.3\% \\
     Llama-3.2-1B-Instruct-Q4 \cite{llama1b} & 773 &
   2250 (1125, 1125) / 1640 (1210, 430) & 27.2\% &
   3375 (1125, 2250) / 2070 (1210, 860) & 38.6\% \\
   Llama-3.2-3B-Instruct-Q4 \cite{llama3b} &  1921 &
      5200 (2600, 2600) / 3530 (2730, 800)   & 32.3\% &
   7800 (2600, 5200) / 4330 (2730, 1600)   & 44.4\% \\
    \bottomrule
  \end{tabular}
  }
  \label{tab:memoryfootprint}
\end{table*}
In this section, we evaluate \name in the context of sharing LLMs between realms.
We assume a setup where multiple realms provide local inference services, each requiring access to a LLM.
We measure the minimum RAM required to run ten typical inference in non-conversational mode, without observing a stall or memory-related error. We define the experiment  under two configurations:
(i) a baseline, where each realm maintains its own instance of the model, and
(ii) model sharing between realms enabled by \name.
We repeat these experiments with two and three realms, using models of different sizes.

\para{Results} \Cref{tab:memoryfootprint} summarizes the results, showing that model sharing between realms reduces the overall system memory footprint by 16.6\% to 44.4\%, depending on the model size and number of realms.
The reduction becomes more pronounced for larger models or when sharing occurs among three realms, demonstrating \names scalability and efficiency in multi-realm deployments.
Notably, these savings represent a conservative \textit{lower bound}. In practical deployments, the benefits of \name could be significantly higher because:
(1) much larger LLMs may be deployed within realms; and
(2) resource sharing can extend beyond LLMs to include read-only user-space binaries and shared packages.
All experiments used the same kernel (43 MB) and filesystem (233 MB).
%%%%%%%%%%%%%%%%%%%%%%%%%%%%%%%%%%%%%%%%%%%%%%%%%%%%%%%%%%%%%%%%%%%%%%
\subsection{Runtime Cost Benchmark}\label{sec:runtimecost}
To evaluate the runtime cost of \name, we conducted an experiment in which we measure inference latency under two configurations. In the baseline setting, the model (GPT-2) is stored in the realm’s private memory. In the second setting, the model is shared with another realm and stored in the CSM. We repeat each experiment using 10 representative queries in non-conversational mode.

\para{Results}
The average inference time is 13.4 seconds in both scenarios. This results shows that executing a model from CSM via the CSM driver achieves \textit{native performance} and that the same physical pages can safely be shared between two realms without incurring any inference-time overhead.
%%%%%%%%%%%%%%%%%%%%%%%%%%%%%%%%%%%%%%%%%%%%%%%%%%%%%%%%%%%%%%%%%%%%%%
\subsection{Memory Mapping Benchmark}\label{sec:memorybench}

\begin{table}[t]
  \centering
  \caption{Average CPU cycles and CPU time per operation}
  \label{tab:memorybench}
  \begin{tabular}{lcc}
    \toprule
    \textbf{Experiment} & \textbf{CPU Cycles} & \makecell{\textbf{CPU Time}\\\textbf{($\mu$s)}} \\
    \midrule

    \multicolumn{3}{l}{\textbf{Operation:} \texttt{RMI\_DATA\_DESTROY}} \\
    \addlinespace[0.2em]
    Vanilla CCA & 5196 & 9.6 \\
    \name       & 5236 & 10.9 \\
    Overhead    & 0.7\%  & 13.5\% \\
    \addlinespace[0.5em]

    \multicolumn{3}{l}{\textbf{Operation:} \texttt{RMI\_DATA\_CREATE\_UNKNOWN}} \\
    \addlinespace[0.2em]
    Vanilla CCA & 3459 & 9.1 \\
    \name       & 3475 & 10.4 \\
    Overhead    & 0.4\%  & 14.2\% \\

    \bottomrule
  \end{tabular}
\end{table}

In this section, we evaluate the impact of \names extensions on memory mapping operations. \name modifies the handlers of \texttt{RMI\_DATA\allowbreak\_CREATE\allowbreak\_UNKNOWN} and \texttt{RMI\_DATA\allowbreak\_DESTROY}, which are invoked when mapping new granules into a realm’s address space and when destroying existing mappings. Although these operations incur one-time costs, they must remain minimal to ensure efficient memory management. To quantify this overhead, we measure the average CPU cycles and execution time, as observed by the hypervisor, required to complete individual \texttt{RMI\_DATA\allowbreak\_CREATE\allowbreak\_UNKNOWN} and \texttt{RMI\_DATA\allowbreak\_DESTROY} operations executed by the RMM. We conduct these measurements under two configurations: the vanilla RMM and the \names RMM.

\para{Results}
Table~\ref{tab:memorybench} presents the average CPU cycles and execution time for the evaluated operations. Overall, \name introduces only a \textit{moderate overhead} compared to the baseline. For \texttt{RMI\_DATA\allowbreak\_DESTROY}, CPU cycles increase from 5196 to 5236 (0.7\%), while execution time increases from 9.6~$\mu$s to 10.9~$\mu$s (13.5\%). For \texttt{RMI\_DATA\allowbreak\_CREATE\allowbreak\_UNKNOWN}, CPU cycles increases from 3459 to 3475 (0.4\%), while execution time increases from 9.1~$\mu$s to 10.4~$\mu$s (14.2\%). Each reported value represents an average over all operations performed during the boot of a realm (approximately 16k \texttt{RMI\_DATA\allowbreak\_DESTROY} operations and 33k \texttt{RMI\_DATA\allowbreak\_CREATE\allowbreak\_UNKNOWN} operations).

\section{Discussion}
In this section, we discuss three topics: why \name remains compatible with (and orthogonal to) future extensions of Arm CCA (\cref{sec:futureextension}); how CSM can be enabled in other confidential computing architectures (\cref{sec:enablingcsminother}); and potential future research directions for \name (\cref{sec:futuredirection}).
\subsection{Arm CCA Future Extensions}\label{sec:futureextension}
Although \name is designed and implemented on top of Arm CCA version~1.0, Arm has recently announced its planned enhancements for CCA version~1.1~\cite{armCCAupdate}. In this section, we describe why \name remains both relevant and fully compatible with the upcoming extensions.

 \para{Planes} Plane extension refers to an architectural extension that enables the decomposition of a realm into multiple EL0\&EL1 execution environments called planes.
Planes are managed by a privileged software component known as the paravisor, which runs in plane~0. The paravisor is responsible for restricting memory access, emulating interrupts, and managing context switching among the other planes. 
Planes were originally proposed to augment a realm with kernel-level runtime services that cannot be provided by the untrusted hypervisor, such as access to a Trusted Platform Module (TPM) \cite{EvolutionofarmCCA}. However, since all planes work in the same realm’s address space, they can also be configured to share confidential memory.
By enabling multi-tenant memory sharing within the realm world, planes can conceptually achieve functionality similar to that of \name. Nevertheless, \name offers  key differences compared to the plane extension, providing a more flexible and decentralized approach.
First, the number of  inter-CVMs memory sharing \name can be dynamically extended at runtime, whereas the number of planes within a realm must be statically defined at boot time.
Second, \name requires no trusted central entity; realms can independently decide where and when to share memory with each other. In contrast, in the plane extension, all planes must trust plane 0 to enforce isolation boundaries and manage sharing, introducing a centralized trust dependency that is unsuitable for scenarios involving multi-party collaboration.

\para{Memory Encryption Context}
The Memory Encryption Context (MEC) extends Arm CCA by introducing hardware-level encryption of memory, providing an additional layer of defense-in-depth \cite{cca,armCCAupdate}. Memory pages tagged with the same encryption context are encrypted using the same encryption key. As proposed by Arm \cite{cca,armCCAupdate}, each realm’s  memory can be protected under a single encryption context. Nevertheless, MEC also allows multiple encryption contexts to coexist within a single realm’s address space \cite{cca}. In other words, a realm vCPU can be configured to access multiple encryption contexts simultaneously. Consequently, a new  encryption context can be assigned to each CSM region, while the memories of the P-realm and C-realm(s) remain encrypted under their own respective contexts. Therefore, \name is fully compatible with MEC hardware extension and MEC's trust model, and it can be extended to leverage the MEC's capabilities.
 
 \para{Device Assignment} This extension enhances  CCA architecture to enable secure assignment of physical devices to realms, a concept previously explored in the literature for GPUs and other generic devices~\cite{bertschi2024devlore,sridhara2024acai,wang2024cage}. Each realm can independently choose whether to allow an off-processor resource, such as an accelerator, to access a region of its address space~\cite{armdeviceenhance,armCCAupdate}. Device assignment extends RSI and RMI and introduces an additional type of metadata granules for realms. We were unable to find any interference between the \name extension of the RMM and the device assignment extension to the RMM.
 %%%%%%%%%%%%%%%%%%%%%%%%%%%%%%%%%%%%%%%%%%%%%%%%%%%%%%%%%%%%%%%%%%%%%%
\subsection{Enabling CSM in Other Architectures} \label{sec:enablingcsminother}
AMD SEV-SNP~\cite{amdsevsnp} and Intel TDX~\cite{inteltdx} are the two major CVM technologies currently deployed by cloud providers, offering security properties comparable to those of realms in Arm CCA.
However, the CISC nature of these architectures makes implementing and evaluating \names functionality significantly more complex-or even infeasible.
This section outlines the limitations of these platforms and discusses potential modifications required to enable CSM-like functionality.

\para{AMD SEV-SNP}
Similar to the GPT in Arm CCA, AMD SEV-SNP introduces a system-wide data structure called the Reverse Map Table (RMP), tracking the ownership and attributes of physical pages.
Unlike the GPT, which only associate pages' entries with a world state, the RMP enforces stricter control by associating pages' entries with CVM's identifier, known as Address Space Identifier (ASID)~\cite{amdsevsnp,SEVSNPfirmware}. AMD's hardware checks the RMP entries at the end of each memory access, granting access to a page only if the ASID of the currently executing guest matches the ASID stored in the corresponding RMP entry. As a result, two CVMs cannot simultaneously access the same confidential page while the hardware-level ASID check is enforced and the RMP only accept one ASID for each page, rendering CSM impossible to implement on SEV-SNP without hardware changes.

\para{Intel TDX}
To the best of our knowledge, Intel TDX~\cite{inteltdx} imposes no inherent hardware limitation that would prevent the implementation of CSM. However, supporting CSM would require modifications to the TDX Module, which has a comparable rule with the RMM in Arm CCA. TDX maintains a system-wide Physical Address Metadata Table (PAMT) that tracks the ownership and state of every physical page.
Whenever a new mapping is established in a CVM’s address space, the TDX Module checks the PAMT to ensure that the corresponding physical pages are not already mapped as protected memory of another CVM.
This restriction could, in principle, be relaxed to permit shared mappings between CVMs that mutually agree to share memory.
The TDX Module also employs Multi-Key Total Memory Encryption (MKTME)~\cite{inteltdxencryption} to encrypt CVM's memory.
MKTME supports memory encryption at page-level granularity, and thus does not impose any inherent restriction on maintaining multiple encrypted regions within a single CVM.
For each CVM, the TDX Module requests a unique encryption key—identified by a Host Key Identifier (HKID)—and embeds it within every page table entry of that CVM, ensuring that all protected memory is encrypted under a single key~\cite{inteltdx,cheng2024intel}.
A CSM-aware TDX implementation would therefore require an additional encryption key dedicated to each new CSM region, with its corresponding page table entries tagged accordingly to preserve compatibility with the existing trust model. We acknowledge that above discussions offer a starting point for adapting CSM in TDX, however, a comprehensive design and concrete implementation merits a separate paper.
%
%However, unlike the open and extensible design of the RMM, the TDX Module must be signed by , limiting the ability of third parties to extend or independently evaluate it with new features such as CSM.

Both AMD's SEV-SNP and Intel's TDX implement intra-CVM isolation mechanisms, \eg the virtual machine privilege levels in AMD.
These isolation mechanisms are similar in nature to planes in CCA, but orthogonal to \name given that they do not allow sharing across different CVMs and follow a strictly hierarchical privilege model that can be limiting~\cite{castes2023creating}.
%%%%%%%%%%%%%%%%%%%%%%%%%%%%%%%%%%%%%%%%%%%%%%%%%%%%%%%%%%%%%%%%%%%%%%
\subsection{Future Direction}\label{sec:futuredirection}
\para{Formal Verification} Arm CCA’s use of formal methods for system design and verification distinguishes it from other confidential computing architectures~\cite{li2023enabling,fox2023verification}.
Although our current work does not provide formal guarantees, an important direction for future research is to apply formal verification techniques to the CSM design within CCA.
Building on recent efforts to formally verify shared-memory mechanisms between enclaves~\cite{lee2022cerberus}, extending the existing formal models of Arm CCA to encompass CSM functionality would represent a significant step toward provable security and functional correctness.

\para{Local Attestation}
As part of establishing the CSM, realms refer to one another solely through attestation-integrated identifiers. In \name, each realm owner is responsible not only for remotely attesting its own realm but also for attesting peer realms with which it intends to associate within the CSM. While remote attestation between a realm owner and its own realm is unavoidable, it is worth exploring whether \name could be redesigned to support local attestation between realms. Such a design—especially in scenarios where no secure channel exists between realm owners and peer realms—could significantly improve system performance while still upholding the security properties expected of attested local interactions.

\para{CVM Signaling over CSM}
\name enables direct memory sharing between CVMs, but the design and integration of signaling mechanisms over this shared substrate remain underexplored.
Developing hypervisor-independent signaling, synchronization, and coordination primitives represents a promising direction for robust inter-CVM interaction.
In particular, integrating CSM-backed signaling with established abstractions such as \textit{virtio} could enhance both the portability and extensibility of inter-CVM communication.
Event-driven synchronization is also feasible through inter-CVM interrupts, which we found can be implemented using standard KVM interfaces—requiring no changes to the hypervisor or the RMM—much like the mechanism employed by \textit{ivshmem}~\cite{ivshmem}.
However, ensuring strong security, isolation, and provenance guarantees for CSM-based signaling remains an open problem and deserves further investigation.

%%%%%%%%%%%%%%%%%%%%%%%%%%%%%%%%%%%%%%%%%%%%%%%%%%%%%%%%%%%%%%%%%%%%%%
\section{Related Works} \label{related-works}
\para{CVM Systems}
Cloud providers already offer CVM instances for a wide range of applications~\cite{CVMazure, googleprivatespace}, with some vendors introducing specialized designs tailored for ML workloads~\cite{awsllm, edgelesssystem}. Recently, new edge-based confidential computing systems have emerged. For instance, Samsung’s Islet~\cite{islet} adopts Arm CCA with a Rust-based RMM, while Aster~\cite{kuhne2024aster} introduces sandboxed realm abstractions to secure Android applications. Similarly, Android’s Virtualization Framework (AVF)~\cite{avfarchitecture} enables VMs that are isolated from the Android kernel layer while providing protected services to Android applications.  

\para{Communication and Memory Sharing} A persistent challenge in current CVM architectures is the performance overhead caused by routing I/O and data exchange through the untrusted hypervisor. It has been shown that hypervisor-mediated services impose significantly higher costs in CVMs than in traditional VMs~\cite{misono2024confidential}. 
Directly sharing memory between two isolated environments can substantially improve the performance of data exchange. This concept has been explored in previous work for conventional VMs~\cite{sreenivasamurthy2019sivshm, ren2016shared} and enclaves~\cite{yu2022elasticlave, lee2022cerberus, li2021confidential}. Specifically, Plug-in Enclave~\cite{li2021confidential} enables read-only shared enclaves for serverless applications, Cerberus~\cite{lee2022cerberus} focuses on the formal verification of memory sharing, and Elasticlave~\cite{yu2022elasticlave} explores sharing models and optimization techniques. Sartakov \etal~\cite{sartakov2022cap} further extend this idea by leveraging CHERI capabilities to allow multiple VM-like compartments to share a single physical address space.

\para{Arm CCA} As Arm CCA gains adoption, research in this area remains limited but is steadily growing. Li \etal~\cite{li2022design} proposed a formal verification methodology for the RMM. Beyond verification, early systems research has explored various extensions to CCA. References~\cite{shen2022soter,sridhara2024acai,wang2024cage,sang2025portal} represent a number of efforts in this space, introducing new features and capabilities to CCA by modifying its trusted components (the RMM and Monitor).
SHELTER~\cite{shen2022soter} leverages the GPC mechanism to support user-space enclaves in the NW. ACAI~\cite{sridhara2024acai} and CAGE~\cite{wang2024cage} address the integration of accelerators into CCA-based systems. Specifically, ACAI enables PCIe accelerators, while CAGE supports the use of integrated GPUs within realms. Portal~\cite{sang2025portal} focuses on secure, high-performance device I/O by enabling direct peripheral access from within a realm on mobile SoCs. 
Other recent work showcases CCA potential for the next-generation of on device ML. GuaranTEE~\cite{siby2024guarantee} is a framework for attestable, privacy-preserving machine learning at the edge. It combines remote attestation and data protection to support collaborative ML inference across devices. Abdollahi \etal~\cite{abdollahi2025early} evaluate the effectiveness of CCA in protecting ML model during inference, validating its applicability for emerging AI workloads.
%%%%%%%%%%%%%%%%%%%%%%%%%%%%%%%%%%%%%%%%%%%%%%%%%%%%%%%%%%%%%%%%%%%%%%
\section{Conclusion}
In this work, we presented \name, the first system that enables CSM, a  hypervisor-protected (confidential) memory which can be shared between multiple CVMs. \name extends Arm CCA firmware with a principled ownership model, explicit access-control rules, and attestation extensions that ensure CSM remains inaccessible to the hypervisor and all unauthorized CVMs, while preserving CCA’s security guarantees for non-CSM memory. 
%The design integrates cleanly across all layers of the CCA stack and retains the hypervisor’s authority over physical memory management, enabling flexible and practical CSM support without requiring hardware changes.
%
\name delivers substantial benefits for communication and data sharing between CVMs. It achieves up to 209$\times$ reduction in CPU cycles compared to encryption-based mechanisms over hypervisor-accessible shared memory, and enables sharing of large data objects such as LLMs with 16.6\%–44.4\% reduction in overall system memory footprint.
\name marks a step toward future compartmentalized and collaborative multi-CVM systems, where each party can protect its proprietary data while efficiently  collaborating with, and providing services to, other parties.

\section*{Acknowledgment}
We thank Jon Crowcroft for his invaluable suggestions that helped improve the paper. We also thank the anonymous reviewers for their insightful comments and suggestions. The research was supported by the UKRI  Open Plus Fellowship (EP/W005271/1, Securing the Next Billion Consumer Devices on the Edge), the Amazon Research Award “Auditable Model Privacy using TEEs”, and the AI Security Institute (AISI) Systemic Safety Grants Programme (UKRI833). 

Professor Kotz was supported by a Royal Society Wolfson Visiting Fellowship and by a collaborative award from the U.S. National Science Foundation (NSF) SaTC Frontiers program under award number 1955805. 

The views and conclusions contained herein are those of the authors and should not be interpreted as necessarily representing the official policies, either expressed or implied, of any sponsor. Any mention of specific companies or products does not imply any endorsement by the authors, by their employers, or by their sponsors. 

Finally note that the authors used  ChatGPT models (GPT-5.1, GPT-5, and GPT-4o) as auxiliary tools for editorial support, exploration of related research, and code debugging. All generated material was examined and verified by the authors, who take full responsibility for the accuracy, integrity, and originality of the paper and the released code.
\bibliographystyle{IEEEtran}
\bibliography{my_bib}

@String{Computing = "Computing" }

@String{Computer = "{IEEE} Computer" }

@misc{avfarchitecture,
  title   = {{AVF architecture}},
 author = {{Android}},
  url     = {https://source.android.com/docs/core/virtualization/architecture\#memory-ownership},
  year    = {2025},
  note    = {Accessed July 2025}
}

@misc{ccasite,
  title   = {{Arm Confidential Compute Architecture}},
  Author   = {{Arm Limited}},
  url     = {https://www.arm.com/architecture/security-features/arm-confidential-compute-architecture},
  year    = {2025},
  note    = {Accessed Feb 2025}
}

@misc{cca,
  title   = {{Introducing Arm Confidential Compute Architecture}},
    Author   = {{Arm Limited}},
  url     = {https://developer.arm.com/documentation/den0125/0300/Overview},
  year    = {2025},
  note    = {Accessed Feb 2025}
}

@misc{ccasoftwarearchitecture,
  title   = {{Arm Confidential Compute Architecture Software Architecture Guide}},
    Author   = {{Arm Limited}},
  url     = {https://developer.arm.com/documentation/den0127/0200/?lang=en},
  year    = {2025},
  note    = {Accessed Feb 2025}
}

@misc{trustzone,
  title   = {{Learn the architecture - TrustZone for AArch64}},
   author  = {{Arm Limited}},
  url     = {https://developer.arm.com/documentation/102418/latest/},
  year    = {2025},
  note    = {Accessed Feb 2025}
}

@misc{armccaenablement,
  title   = {{Arm Confidential Compute Architecture open-source enablement}},
  organization  = {{The Linux Foundation}},
  url     = {https://confidentialcomputing.io/webinars/arm-confidential-compute-architecture-open-source-enablement/},
  year    = {2025},
  note    = {Accessed Feb 2025}
}

@misc{fvp,
  title   = {{Fast Models Fixed Virtual Platforms (FVP) Reference Guide}},
   Author  = {{Arm Limited}},
  url     = {https://developer.arm.com/Tools\%20and\%20Software/Fixed\%20Virtual\%20Platforms},
  year    = {2025},
  note    = {Accessed Feb 2025}
}

@misc{armccasecurity,
  title   = {{Arm CCA Security Model 1.0}},
   Author  = {{Arm Limited}},
  url     = {https://developer.arm.com/documentation/DEN0096/latest},
  year    = {2025},
  note    = {Accessed Feb 2025}
}

@misc{FastModel,
  title   = {{Fast Models Reference Guide}},
   Author  = {{Arm Limited}},
  url     = {https://developer.arm.com/Tools\%20and\%20Software/Fixed\%20Virtual\%20Platforms},
  year    = {2025},
  note    = {Accessed Feb 2025}
}

@misc{rmmspec,
  title   = {{Realm Management Monitor Specification}},
    Author  = {{Arm Limited}},
  url     = {https://developer.arm.com/documentation/den0137/1-0eac5/?lang=en},
  year    = {2025},
  note    = {Accessed Feb 2025}
}

@misc{rmmlock,
  title   = {{RMM Locking Guidelines}},
  author  = { },
  url     = {https://tf-rmm.readthedocs.io/en/latest/design/locking.html},
  year    = {2025},
  note    = {Accessed April 2025}
}

@misc{ivshmem,
  title   = {{Inter-VM Shared Memory device}},
  Author  = {{QEMU Project}},
  url     = {https://www.qemu.org/docs/master/system/devices/ivshmem.html},
  year    = {2023}
}

@misc{armdeviceenhance,
  title   = {{Arm A-Profile Architecture Developments 2022}},
  author  = {Martin Weidmann},
  url     = {https://community.arm.com/arm-community-blogs/b/architectures-and-processors-blog/posts/arm-a-profile-architecture-2022l},
  year    = {2022}
}

@misc{TF-A,
  title   = {{TF-A}},
  author = {trusted firmware},
  url     = {https://www.trustedfirmware.org/projects/tf-a},
  year    = {2025},
  note    = {Accessed Feb 2025}
}

@misc{RMM,
  title   = {{TF-RMM}},
  Author  = {{TrustedFirmware}},
  url     = {https://www.trustedfirmware.org/projects/tf-rmm},
  year    = {2025},
  note    = {Accessed Feb 2025}
}

@misc{linux-cca,
  title   = {linux-cca},
   Author  = {{Arm Limited}},
  url     = {https://gitlab.arm.com/linux-arm/linux-cca/-/commit/fad35572db},
  year    = {2025},
  note    = {Accessed Feb 2025}
}

@misc{llamacpp,
  title   = {{llama.cpp}},
  author  = {Georgi Gerganov},
  url     = {https://github.com/ggerganov/llama.cpp},
  year    = {2023},
  note    = {Accessed Feb 2025}
}

@misc{gpt2,
  title   = {{openai-community/GPT2}},
  author  = {{Hugging Face}},
  url     = {https://huggingface.co/openai-community/gpt2},
  year = {2019},
  note    = {Accessed Feb 2025}
}

@misc{gpt2-medium,
  title   = {{openai-community/gpt2-medium}},
  author  = {{Hugging Face}},
    year = {2019},
  url     = {https://huggingface.co/openai-community/gpt2-medium},
  note    = {Accessed Feb 2025}
}

@misc{llama1b,
  title   = {{bartowski/Llama-3.2-1B-Instruct-GGUF}},
  Author  = {{Hugging Face}},
  url     = {https://huggingface.co/bartowski/Llama-3.2-1B-Instruct-GGUF},
  note    = {Accessed Feb 2025}
}

@misc{llama3b,
  title   = {{bartowski/Llama-3.2-3B-Instruct-GGUF}},
  Author  = {{Hugging Face}},
  url     = {https://huggingface.co/bartowski/Llama-3.2-3B-Instruct-GGUF},
  note    = {Accessed Feb 2025}
}

@article{fox2023verification,
  title     = {{A Verification Methodology for the Arm{\textregistered} Confidential Computing Architecture: From a Secure Specification to Safe Implementations}},
  author    = {Fox, Anthony CJ and Stockwell, Gareth and Xiong, Shale and Becker, Hanno and Mulligan, Dominic P and Petri, Gustavo and Chong, Nathan},
  journal   = {Proceedings of the ACM on Programming Languages},
  volume    = {7},
  number    = {OOPSLA1},
  pages     = {376--405},
  year      = {2023},
  publisher = {ACM}
}

@article{li2023enabling,
  title={{Enabling Realms with the Arm Confidential Compute Architecture}},
  author={Li, Xupeng and Li, Xuheng and Dall, Christoffer and Gu, Ronghui and Nieh, Jason and Sait, Yousuf and Stockwell, Gareth and Knight, Mark and Garcia-Tobin, Charles},
  year={2023}
}

@inproceedings{li2022design,
  title     = {{Design and verification of the Arm confidential compute architecture}},
  author    = {Li, Xupeng and Li, Xuheng and Dall, Christoffer and Gu, Ronghui and Nieh, Jason and Sait, Yousuf and Stockwell, Gareth},
  booktitle = {16th USENIX Symposium on Operating Systems Design and Implementation (OSDI 22)},
  pages     = {465--484},
  year      = {2022}
}

@article{sardar2023sok,
  title   = {{SoK: Attestation in confidential computing}},
  author  = {Sardar, M and Fossati, Thomas and Frost, Simon},
  journal = {ResearchGate pre-print},
  year    = {2023}
}

@inproceedings{shen2022soter,
  title     = {{SOTER: Guarding Black-box Inference for General Neural Networks at the Edge}},
  author    = {Shen, Tianxiang and Qi, Ji and Jiang, Jianyu and Wang, Xian and Wen, Siyuan and Chen, Xusheng and Zhao, Shixiong and Wang, Sen and Chen, Li and Luo, Xiapu and others},
  booktitle = {2022 USENIX Annual Technical Conference (USENIX ATC 22)},
  pages     = {723--738},
  year      = {2022}
}

@inproceedings{sridhara2024acai,
  title     = {{ACAI: Extending Arm Confidential Computing Architecture Protection from CPUs to Accelerators}},
  author    = {Sridhara, Supraja and Bertschi, Andrin and Schl{\"u}ter, Benedict and Kuhne, Mark and Aliberti, Fabio and Shinde, Shweta},
  booktitle = {33rd USENIX Security Symposium (USENIX Security’24)},
  year      = {2024}
}

@inproceedings{zhang2023shelter,
  title     = {{SHELTER: Extending Arm CCA with Isolation in User Space}},
  author    = {Zhang, Yiming and Hu, Yuxin and Ning, Zhenyu and Zhang, Fengwei and Luo, Xiapu and Huang, Haoyang and Yan, Shoumeng and He, Zhengyu},
  booktitle = {32nd USENIX Security Symposium (USENIX Security’23)},
  year      = {2023}
}

@inproceedings{zhang2024tsdp,
  title     = {{No Privacy Left Outside: On the (In-) Security of TEE-Shielded DNN Partition for On-Device ML}},
  author    = {Zhang, Ziqi and Gong, Chen and Cai, Yifeng and Yuan, Yuanyuan and Liu, Bingyan and Li, Ding and Guo, Yao and Chen, Xiangqun},
  booktitle = {2024 IEEE Symposium on Security and Privacy (SP)},
  pages     = {52--52},
  year      = {2024},
  organization = {IEEE Computer Society}
}

@inproceedings{wang2024cage,
  title     = {{CAGE: Complementing Arm CCA with GPU Extensions}},
  author    = {Wang, Chenxu and Zhang, Fengwei and Deng, Yunjie and Leach, Kevin and Cao, Jiannong and Ning, Zhenyu and Yan, Shoumeng and He, Zhengyu},
  booktitle = {Network and Distributed System Security (NDSS) Symposium},
  year      = {2024}
}

@article{bertschi2024devlore,
  title   = {{Devlore: Extending Arm CCA to Integrated Devices A Journey Beyond Memory to Interrupt Isolation}},
  author  = {Bertschi, Andrin and Sridhara, Supraja and Groschupp, Friederike and Kuhne, Mark and Schl{\"u}ter, Benedict and Thorens, Cl{\'e}ment and Dutly, Nicolas and Capkun, Srdjan and Shinde, Shweta},
  journal = {arXiv preprint arXiv:2408.05835},
  year    = {2024}
}

@article{kuhne2024aster,
  title   = {{Aster: Fixing the Android TEE ecosystem with Arm CCA}},
  author  = {Kuhne, Mark and Sridhara, Supraja and Bertschi, Andrin and Dutly, Nicolas and Capkun, Srdjan and Shinde, Shweta},
  journal = {arXiv preprint arXiv:2407.16694},
  year    = {2024}
}

@inproceedings{siby2024guarantee,
  title     = {{GuaranTEE: Towards Attestable and Private ML with CCA}},
  author    = {Siby, Sandra and Abdollahi, Sina and Maheri, Mohammad and Kogias, Marios and Haddadi, Hamed},
  booktitle = {Proceedings of the 4th Workshop on Machine Learning and Systems},
  pages     = {1--9},
  year      = {2024}
}

@misc{QEMUlinaro,
  title   = {{qemu}},
  Author  = {{Linaro}},
  url     = {https://git.codelinaro.org/linaro/dcap/qemu},
  year    = {2025},
  note    = {Accessed Feb 2025}
}

@misc{islet,
  title   = {Islet},
  Author  = {{Samsung}},
  url     = {https://github.com/islet-project/islet},
  year    = {2025},
  note    = {Accessed Feb 2025}
}

@misc{kvmtool-cca,
  title   = {{kvmtool-cca}},
  Author   = {},
  url     = {https://gitlab.arm.com/linux-arm/kvmtool-cca/-/tree/cca/v3?ref_type=heads},
  year    = {2025},
  note    = {Accessed Feb 2025}
}

@inproceedings{castes2023creating,
  title     = {{Creating trust by abolishing hierarchies}},
  author    = {Castes, Charly and Ghosn, Adrien and Kalani, Neelu S and Qian, Yuchen and Kogias, Marios and Payer, Mathias and Bugnion, Edouard},
  booktitle = {Proceedings of the 19th Workshop on Hot Topics in Operating Systems},
  pages     = {231--238},
  year      = {2023}
}

@misc{edgelesssystem,
  Author       = {{Edgeless Systems GmbH}},
  year         = {2025},
  month        = mar,
  title        = {{The always encrypted AI service}},
  note = {March 7, 2025},
  url          = {https://www.privatemode.ai/}
}

@misc{applepcc,
  author       = {{Apple Security Engineering and Architecture (SEAR)} and {User Privacy} and {Core Operating Systems (Core OS)} and {Services Engineering (ASE)} and {Machine Learning and AI (AIML)}},
  year         = {2024},
  title        = {Private Cloud Compute: A new frontier for AI privacy in the cloud},
  note = {March 9, 2025},
  url          = {https://security.apple.com/blog/private-cloud-compute/}
}

@misc{armCCAupdate,
  author       = {Linaro},
  year         = {2024},
  title        = {MAD24-410 Arm Confidential Compute Architecture open-source enablement update},
  lastaccessed = {March 9, 2025},
  url          = {https://resources.linaro.org/en/resource/rEjhEezEvnNMC3LALzUTrr}
}

@misc{googleprivatespace,
 author       = {{Google Cloud}},
  year         = {},
  title        = {{Confidential Space security overview}},
  lastaccessed = {June 7, 2025},
  url          = {https://cloud.google.com/docs/security/confidential-space#:~:text=,resource\%20is\%20protected\%20by\%20an}
}

@misc{awsllm,
  author       = {Chris Renzo and Liv d’Aliberti and Justin Miles and Joe Kovba},
  year         = {2024},
  title        = {{Large language model inference over confidential data using AWS Nitro Enclaves}},
  lastaccessed = {June 10, 2025},
  url          = {https://aws.amazon.com/blogs/machine-learning/large-language-model-inference-over-confidential-data-using-aws-nitro-enclaves/}
}

@misc{amdsevsnp,
  author       = {AMD},
  year         = {},
  title        = {{AMD SEV-SNP: Strengthening VM Isolation with Integrity Protection and More}},
  lastaccessed = {June 7, 2025},
 url = {https://www.amd.com/content/dam/amd/en/documents/epyc-business-docs/white-papers/SEV-SNP-strengthening-vm-isolation-with-integrity-protection-and-more.pdf}
}

@misc{inteltdx,
 author   = {{Intel®}},
  year         = {2025},
  title        = {{Intel® Trust Domain Extensions (Intel TDX)}},
  lastaccessed = {June 7, 2025},
  url          = {{https://cdrdv2.intel.com/v1/dl/getContent/690419}}
}

@misc{inteltdxencryption,
 author   = {{Intel®}},
  year         = {2025},
  title        = {{Intel® Architecture Memory Encryption Technologies Specification}},
  lastaccessed = {September 7, 2025},
  url          = {{https://www.intel.com/content/www/us/en/content-details/679154/intel-architecture-memory-encryption-technologies-specification.html}}
}

@article{cheng2024intel,
  title={Intel tdx demystified: A top-down approach},
  author={Cheng, Pau-Chen and Ozga, Wojciech and Valdez, Enriquillo and Ahmed, Salman and Gu, Zhongshu and Jamjoom, Hani and Franke, Hubertus and Bottomley, James},
  journal={ACM Computing Surveys},
  volume={56},
  number={9},
  pages={1--33},
  year={2024},
  publisher={ACM New York, NY}
}

@misc{intelsgx,
 author    = {{Intel®}},
  year         = {2025},
  title        = {{Intel Software Guard Extensions}},
  lastaccessed = {June 7, 2025},
  url          = {{https://www.intel.com/content/www/us/en/developer/tools/software-guard-extensions/overview.html}}
}

@misc{SEVSNPfirmware,
  year         = {2025},
  title        = {{SEV Secure Nested Paging Firmware ABI Specification}},
  lastaccessed = {June 7, 2025},
  url          = {https://docs.amd.com/v/u/en-US/56860}
}

@misc{EvolutionofarmCCA,
  title        = {Evolution of the Arm Confidential Compute Architecture by G. Stockwell, N. Sample \& P. Howard | OC3},
  lastaccessed = {April 15, 2025},
  url          = {https://www.youtube.com/watch?v=1AsvIt7bSLY&t=2086s}
}

@misc{nvidiaedgeless,
  organization       = {{NVIDIA}},
  year         = {2024},
  title        = {{Advancing Security for Large Language Models with NVIDIA GPUs and Edgeless Systems}},
  lastaccessed = {May 9, 2025},
  url          = {https://developer.nvidia.com/blog/advancing-security-for-large-language-models-with-nvidia-gpus-and-edgeless-systems/}
}

@misc{CVMazure,
   author        = {{Microsoft Azure}},
  year         = {2023},
  title        = {{Confidential VMs on Azure}},
  lastaccessed = {June 9, 2025},
  url          = {https://techcommunity.microsoft.com/blog/windowsosplatform/confidential-vms-on-azure/3836282}
}

@article{dall2014kvm,
  title     = {{KVM/ARM: the design and implementation of the Linux ARM hypervisor}},
  author    = {Dall, Christoffer and Nieh, Jason},
  journal   = {ACM Sigplan Notices},
  volume    = {49},
  number    = {4},
  pages     = {333--348},
  year      = {2014},
  publisher = {ACM}
}

@misc{kvm,
 author = {{Wikipedia}},
  year         = {2025},
  title        = {{Kernel-based Virtual Machine}},
  lastaccessed = {June 7, 2025},
  url          = {{https://en.wikipedia.org/wiki/Kernel-based_Virtual_Machine}}
}

@inproceedings{lefeuvre2023towards,
  title     = {{Towards (really) safe and fast confidential I/O}},
  author    = {Lefeuvre, Hugo and Chisnall, David and Kogias, Marios and Olivier, Pierre},
  booktitle = {Proceedings of the 19th Workshop on Hot Topics in Operating Systems},
  pages     = {214--222},
  year      = {2023}
}

@article{sreenivasamurthy2019sivshm,
  title   = {{SIVSHM: Secure inter-vm shared memory}},
  author  = {Sreenivasamurthy, Shesha and Miller, Ethan},
  journal = {arXiv preprint arXiv:1909.10377},
  year    = {2019}
}

@inproceedings{li2023bifrost,
  title     = {{Bifrost: Analysis and optimization of network $\{$I/O$\}$ tax in confidential virtual machines}},
  author    = {Li, Dingji and Mi, Zeyu and Ji, Chenhui and Tan, Yifan and Zang, Binyu and Guan, Haibing and Chen, Haibo},
  booktitle = {2023 USENIX Annual Technical Conference (USENIX ATC 23)},
  pages     = {1--15},
  year      = {2023}
}

@article{misono2024confidential,
  title     = {{Confidential VMs Explained: An Empirical Analysis of AMD SEV-SNP and Intel TDX}},
  author    = {Misono, Masanori and Stavrakakis, Dimitrios and Santos, Nuno and Bhatotia, Pramod},
  journal   = {Proceedings of the ACM on Measurement and Analysis of Computing Systems},
  volume    = {8},
  number    = {3},
  pages     = {1--42},
  year      = {2024},
  publisher = {ACM}
}

@article{chen2024protecting,
  title   = {{Protecting Confidentiality, Privacy and Integrity in Collaborative Learning}},
  author  = {Chen, Dong and Dethise, Alice and Akkus, Istemi Ekin and Rimac, Ivica and Satzke, Klaus and Koskela, Antti and Canini, Marco and Wang, Wei and Chen, Ruichuan},
  journal = {arXiv preprint arXiv:2412.08534},
  year    = {2024}
}

@inproceedings{lee2022cerberus,
  title     = {{Cerberus: A formal approach to secure and efficient enclave memory sharing}},
  author    = {Lee, Dayeol and Cheang, Kevin and Thomas, Alexander and Lu, Catherine and Gaddamadugu, Pranav and Vahldiek-Oberwagner, Anjo and Vij, Mona and Song, Dawn and Seshia, Sanjit A and Asanovic, Krste},
  booktitle = {Proceedings of the 2022 ACM SIGSAC Conference on Computer and Communications Security},
  pages     = {1871--1885},
  year      = {2022}
}

@article{abdollahi2025early,
  title   = {{An Early Experience with Confidential Computing Architecture for On-Device Model Protection}},
  author  = {Abdollahi, Sina and Maheri, Mohammad and Siby, Sandra and Kogias, Marios and Haddadi, Hamed},
  journal = {arXiv preprint arXiv:2504.08508},
  year    = {2025}
}

@inproceedings{yu2022elasticlave,
  title     = {{Elasticlave: An efficient memory model for enclaves}},
  author    = {Yu, Jason Zhijingcheng and Shinde, Shweta and Carlson, Trevor E and Saxena, Prateek},
  booktitle = {31st USENIX Security Symposium (USENIX Security 22)},
  pages     = {4111--4128},
  year      = {2022}
}

@article{bertschi2025opencca,
  title   = {{OpenCCA: An Open Framework to Enable Arm CCA Research}},
  author  = {Bertschi, Andrin and Shinde, Shweta},
  journal = {arXiv preprint arXiv:2506.05129},
  year    = {2025}
}

@inproceedings{moon2025asgard,
  title={{ASGARD: Protecting On-Device Deep Neural Networks with Virtualization-Based Trusted Execution Environments}},
  author={Moon, Myungsuk and Kim, Minhee and Jung, Joonkyo and Song, Dokyung},
  booktitle={Proceedings 2025 Network and Distributed System Security Symposium},
  year={2025}
}

@inproceedings{li2021confidential,
  title={{Confidential serverless made efficient with plug-in enclaves}},
  author={Li, Mingyu and Xia, Yubin and Chen, Haibo},
  booktitle={2021 ACM/IEEE 48th Annual International Symposium on Computer Architecture (ISCA)},
  pages={306--318},
  year={2021},
  organization={IEEE}
}

@inproceedings{sang2025portal,
  title={{PORTAL: Fast and Secure Device Access with Arm CCA for Modern Arm Mobile System-on-Chips (SoCs)}},
  author={Sang, Fan and Lee, Jaehyuk and Zhang, Xiaokuan and Kim, Taesoo},
  booktitle={2025 IEEE Symposium on Security and Privacy (SP)},
  pages={4099--4116},
  year={2025},
  organization={IEEE}
}

@inproceedings{sartakov2022cap,
  title={{CAP-VMs: Capability-Based Isolation and Sharing in the Cloud}},
  author={Sartakov, Vasily A and Vilanova, Llu{\'\i}s and Eyers, David and Shinagawa, Takahiro and Pietzuch, Peter},
  booktitle={16th USENIX Symposium on Operating Systems Design and Implementation (OSDI 22)},
  pages={597--612},
  year={2022}
}

@article{ren2016shared,
  title={{Shared-memory optimizations for inter-virtual-machine communication}},
  author={Ren, Yi and Liu, Ling and Zhang, Qi and Wu, Qingbo and Guan, Jianbo and Kong, Jinzhu and Dai, Huadong and Shao, Lisong},
  journal={ACM Computing Surveys (CSUR)},
  volume={48},
  number={4},
  pages={1--42},
  year={2016},
  publisher={ACM New York, NY, USA}
}

@article{wu2024isolategpt,
  title={{Isolategpt: An execution isolation architecture for llm-based agentic systems}},
  author={Wu, Yuhao and Roesner, Franziska and Kohno, Tadayoshi and Zhang, Ning and Iqbal, Umar},
  journal={arXiv preprint arXiv:2403.04960},
  year={2024}
}

@misc{vmcoreandroid,
  author = {{Android Developers}},
  title        = {{Virtual Machine as a core Android Primitive}},
  lastaccessed = {April 14, 2025},
  url          = {https://android-developers.googleblog.com/2023/12/virtual-machines-as-core-android-primitive.html}
}

@misc{ROCK5B,
  author = {{Radxa}},
  title        = {{ROCK 5B}},
  lastaccessed = {April 14, 2025},
  url          = {{https://radxa.com/products/rock5/5b/}}
}

@misc{ROCK5Bplus,
  author = {{Radxa}},
  title        = {{ROCK 5B+}},
  lastaccessed = {April 14, 2025},
  url          = {{https://radxa.com/products/rock5/5bp}}
}

\appendices

% \section{Data Availability}
% All data and code required to reproduce our evaluations will be open-sourced at the time of acceptance. 

\vspace{12pt}
\end{document}